\documentclass[
nofootinbib,
reprint,
superscriptaddress, prl,
]{revtex4-1}

\usepackage{physics}
\usepackage{bm}
\usepackage[T1]{fontenc}
\usepackage{amsmath,amsfonts,bbm, graphicx, color}
\usepackage{amssymb}
\usepackage{appendix}
\usepackage{mathtools}
\usepackage{dsfont}
\usepackage{float}
\usepackage{relsize}
\usepackage{xcolor}
\usepackage{scalerel}
\usepackage{tikz}
\usetikzlibrary{svg.path}
\usepackage{soul}

\definecolor{orcidlogocol}{HTML}{A6CE39}
\tikzset{
  orcidlogo/.pic={
    \fill[orcidlogocol] svg{M256,128c0,70.7-57.3,128-128,128C57.3,256,0,198.7,0,128C0,57.3,57.3,0,128,0C198.7,0,256,57.3,256,128z};
    \fill[white] svg{M86.3,186.2H70.9V79.1h15.4v48.4V186.2z}
                 svg{M108.9,79.1h41.6c39.6,0,57,28.3,57,53.6c0,27.5-21.5,53.6-56.8,53.6h-41.8V79.1z M124.3,172.4h24.5c34.9,0,42.9-26.5,42.9-39.7c0-21.5-13.7-39.7-43.7-39.7h-23.7V172.4z}
                 svg{M88.7,56.8c0,5.5-4.5,10.1-10.1,10.1c-5.6,0-10.1-4.6-10.1-10.1c0-5.6,4.5-10.1,10.1-10.1C84.2,46.7,88.7,51.3,88.7,56.8z};
  }
}

\newcommand\orcid[1]{\href{https://orcid.org/#1}{\mbox{\scalerel*{
\begin{tikzpicture}[yscale=-1,transform shape]
\pic{orcidlogo};
\end{tikzpicture}
}{|}}}}
\usepackage[colorlinks]{hyperref}

\hypersetup{
	bookmarksnumbered,
	pdfstartview={FitH},
	citecolor={blue},
	linkcolor={darkblue},
	urlcolor={blue},
	pdfpagemode={UseOutlines}}
\definecolor{darkgreen}{RGB}{20,100,20}
\definecolor{darkblue}{RGB}{0,0,130}
\definecolor{darkred}{rgb}{.8,0,0}

\newcommand{\updownarrows}{\uparrow\mathrel{\mspace{-1mu}}\downarrow}
\newcommand{\downuparrows}{\downarrow\mathrel{\mspace{-1mu}}\uparrow}

\begin{document}

\title{Indefinite temporal order without gravity}

\author{Kacper~D\k{e}bski\orcid{0000-0002-8865-9066}}
\email{kdebski@fuw.edu.pl}
\affiliation{Institute of Theoretical Physics, University of Warsaw, Pasteura 5, 02-093 Warsaw, Poland}

\author{Magdalena Zych}
\email{m.zych@uq.edu.au}

\author{Fabio Costa}
\email{f.ccosta@uq.edu.au}
\affiliation{Australian Research Council Centre for Engineered Quantum Systems,
School of Mathematics and Physics, The University of Queensland, St Lucia, QLD 4072, Australia}

\author{Andrzej~Dragan\orcid{0000-0002-5254-710X}}
\email{dragan@fuw.edu.pl}
\affiliation{Institute of Theoretical Physics, University of Warsaw, Pasteura 5, 02-093 Warsaw, Poland}
\affiliation{Centre for Quantum Technologies, National University of Singapore, 3 Science Drive 2, 117543 Singapore, Singapore}

\date{\today}
 
\begin{abstract}
According to the general theory of relativity, time can flow at different rates depending on the configuration of massive objects, affecting the temporal order of events. Recent research has shown that, combined with quantum theory, this gravitational effect can result in events with an indefinite temporal order, which can be tested through the violation of Bell-type inequalities. According to Einstein, we shall assume physical equivalence of a uniform gravitational field and a corresponding acceleration of a reference system. Here we construct a non-gravitational scenario where accelerating particles interacting with optical cavities result in a violation of the temporal Bell inequalities analogous to the gravitational case. However, we find that the inequalities can also be violated by time-like events, exposing an ambiguity in their use as a theory-independent test of indefinite temporal order. 
\end{abstract}

\maketitle

\section*{Introduction}
Quantum theory and general relativity rest on very different foundations. In quantum theory, systems do not posses definite physical properties independent of their measurement; however, processes take place on a fixed background spacetime, where the causal relations between events---space-like or time-like---are defined independently of any operation or physical process. On the other hand, in general relativity, causal relations are not fixed a priori, as the geometry depends on the configuration of mass-energy. Therefore, it is expected that a unification of the two theories should result in non-classical, or \emph{indefinite}, causal structure \cite{Butterfield2001,hardy2007towards}.

However, the physical meaning of such an indefinite causal structure is not clear. Most approaches to quantum gravity attempt to establish a complete theoretical framework \cite{oriti2009, kiefer2012quantum}, but do not offer a direct physical interpretation of non-classical causal relations. 

Recent work has proposed a Gedankenexperiment to directly pinpoint the physical meaning of the non-classicality of temporal order, independently of the details of a full quantum theory of gravity \cite{Bell4time}. The idea is that a superposition of mass configurations will produce a corresponding non-classical time dilation on any system one might use as ``clock'' to identify spacetime events, for example, producing effectively a superposition of orderings of time-like events.
By choosing an appropriate protocol, one can perform a task---the violation of a Bell inequality---that would be impossible if the events were ordered according to a classical causal structure. This would provide a theory independent certification of the non-classicality of causal structure emerging from the incorporation of the superposition principle into general relativity.

As the proposal in Ref.~\cite{Bell4time} relies on time dilation, it is legitimate to ask whether a similar experiment would be possible in a purely special-relativistic setting, where the  back-action on spacetime of the involved matter is negligible but where a superposition of the clock states of motion causes  the corresponding non-classical time dilation. Furthermore, a flat spacetime version of the argument would allow in principle a complete/fundamental  description of a potential practical implementation and thus provide insights into the requirements entering the original, gravitational, argument.
Special relativistic implementation of the experiment might also be less demanding and provide a possible outlook for feasible laboratory implementations.

Here we present such an in principle complete scheme where accelerating particles 
interact with quantum fields according to their own internal clock degrees of freedom (DoFs). By letting the particles evolve in superposition along appropriately accelerating trajectories, the particle-field interaction events reproduce the ``entanglement of temporal order'' found in the gravitational case. Further measurements on the particles and fields can thus violate the ``Bell inequalities for temporal order'' presented in Ref.~\cite{Bell4time}.

Surprisingly, we find that a violation of the Bell inequalities persists when the entangled  events are space-like, challenging the interpretation that the protocol demonstrates a non-classical temporal order of time-like events. We find that this is due to the failure of one of the assumptions behind the original protocol---namely that the superposed alternatives only differ in the event order, while all local evolutions are trivial.
We argue that the failure of this assumption is ubiquitous and would be present in a generic  dynamical context, including a gravitational implementation of the protocol. We also propose an interpretation of the Bell inequality violation in scenarios where non-classicality of temporal order cannot possibly explain the results (e.g. for spacelike separated events, as mentioned above).

Our results raise a fundamental question, whether it is possible to formulate an operational scenario that can unambiguously discriminate the non-classicality of the causal structure of spacetime from dynamical effects, necessarily present in its physical implementations, or other laboratory implementations of  quantum causal structures \cite{Procopio2015, Rubinoe1602589, rubino2017experimental, Goswami2018, guo2018experimental, Wei2019, taddei2020experimental}. We present a discussion on what extensions to the original protocol might be required in order to achieve that.

Throughout this work we use natural units, for which we have $\hbar=c=1$.

\subsection{Gravitational implementation of an indefinite causal order}

For later reference. we will briefly review the salient aspects of the protocol introduced in Ref.~\cite{Bell4time}. The goal of the protocol is to realise four events, $A_1$, $B_1$, $A_2$, $B_2$, whose pair-wise order is `entangled': $A_1$ is in the causal past/past lightcone of $B_1$, denoted $A_1 \prec B_1$, when $A_2 \prec B_2$, and $A_1$ is in the causal future/future lightcone of $B_1$, denoted $A_1 \succ B_1$ when $A_2 \succ B_2$; 
the full scenario is arranged such that this entanglement leads to correlations that do not admit a classical explanation, which is formalised in analogy to Bell-like scenarios for local classical properties. Crucially, an ``event'' is here understood operationally as ``something that happens at a particular time and place'' and thus is defined by some physical reference system. In the protocol, the reference systems are four clocks, $a_1$, $b_1$, $a_2$, $b_2$, while the events are associated with quantum operations performed on some additional system when the corresponding clock reaches a specified proper time: $A_1$ takes place (an operation is performed) when $a_1$ reaches time $\tau^*$, $B_1$ takes place when $b_1$ reaches time $\tau^*$, and so on.

\begin{figure}[h!tbp]
\centering
    \includegraphics[width=1\linewidth]{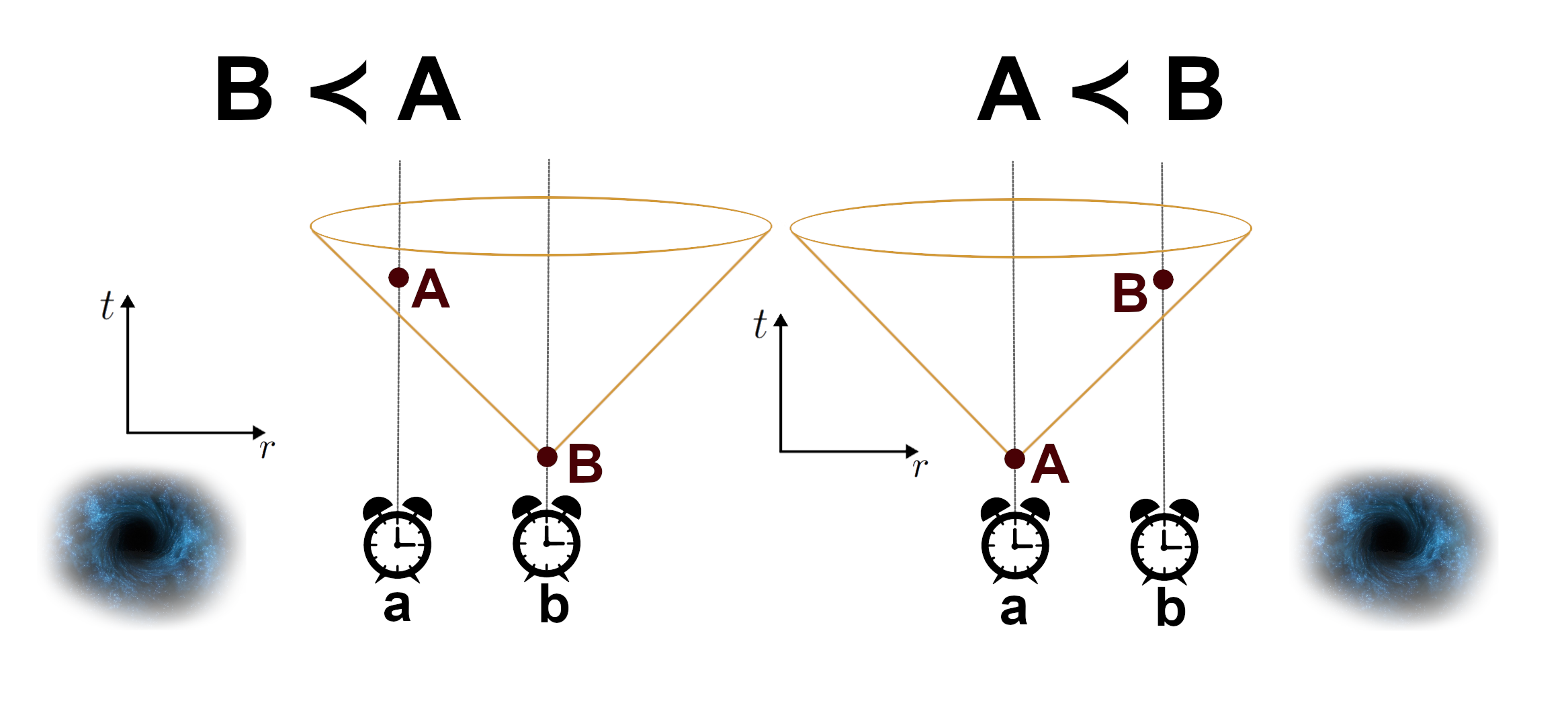}
    \caption{\label{figuregravitational} {\bf Position of a mass as a control of time order.} Two identical clocks, $a,b$ are synchronised (with any source-mass sufficiently far away). Events $A, B$ are defined as the location and fixed proper time $\tau^*$ of the corresponding clock. In the absence of any source mass, events $A, B$ are space-like. However, if a massive object is initially (just after synchronisation) placed closer to clock $a$ than to $b$, event $A$ can be in the future light cone of $B$ (for sufficiently large $\tau^*$), denoted $B \prec A$. Analogously, for the mass closer to $b$,  event $A$ can be up in the past light cone of $B$,  $A \prec B$. }
\end{figure}

In flat spacetime, if the clocks are initially synchronised (in an arbitrary chosen reference frame), all events are space-like separated. However, introducing a massive body closer to some clocks than others causes a differential time dilation, which can `push' some events into the future light-cone of other events. Thereby, the position of the mass provides a control of the time order of events, see Fig.~\ref{figuregravitational}. 
If the control mass is prepared in a semiclassical configuration denoted $K_A$, event $A_j$, $j=1,2$  is in the past of $B_j$, while for  a different configuration, $K_B$, event $B_j$ is in the past of $A_j$. Thus, $A_j$ is time-like from $B_j$, for each mass configuration, but their order is interchanged.  Moreover, for both mass configurations the pair $A_1$, $B_1$ is space-like from $A_2$, $B_2$. 

The full protocol features also two of the above mentioned additional systems on which the operations are performed, $S_1$ and $S_2$, referred to as 'targets'. The operations at events $A_1$ and $B_1$ are applied only on the target system $S_1$, while those at $A_2$ and $B_2$ are applied only on $S_2$. The two target systems are initially  in a product state, $\ket{\psi}^{S_1}\ket{\psi}^{S_2}$, and the considered operations are unitaries: $\hat{U}^{A_1}$ is applied on $S_1$ at event $A_1$, etc. 
By preparing the control mass in a superposition state, 
$\ket{K} = \frac{1}{\sqrt{2}}\left(\ket{K_A}+\ket{K_B}\right)$, one obtains the final state
\begin{multline} \label{graviswitch}
\ket{\Psi^{\textrm{fin}}} =
\frac{1}{\sqrt{2}}
\left(
\ket{K_A}\hat{U}^{B_1}\hat{U}^{A_1}\ket{\psi}^{S_1} \hat{U}^{B_2}\hat{U}^{A_2}\ket{\psi}^{S_2} \right.\\
+ \left. \ket{K_B}\hat{U}^{A_1}\hat{U}^{B_1}\ket{\psi}^{S_1} \hat{U}^{A_2}\hat{U}^{B_2}\ket{\psi}^{S_2}
\right).
\end{multline}

Next, the control mass is measured in the basis $\ket{\pm} = \frac{1}{\sqrt{2}}\left(\ket{K_A}\pm\ket{K_B}\right)$, leaving the target system in
\begin{multline}
\ket{\Psi^{\textrm{post}}} =
\frac{1}{\sqrt{2}}
\left(
\ket{\psi_A}^{S_1} \ket{\psi_A}^{S_2} \pm \ket{\psi_B}^{S_1} \ket{\psi_B}^{S_2}
\right),
\end{multline}
where $\ket{\psi_A} = \hat{U}^{B}\hat{U}^{A}\ket{\psi}$, $\ket{\psi_B} = \hat{U}^{A}\hat{U}^{B}\ket{\psi}$, and we are using the same unitaries in the two `wings': $\hat{U}^{A_1} = \hat{U}^{A_2} \equiv \hat{U}^{A}$, $\hat{U}^{B_1} = \hat{U}^{B_2} \equiv \hat{U}^{B}$.

In general,  $\ket{\Psi^{\textrm{post}}}$ is an entangled state, unless $\ket{\psi_A} = \ket{\psi_B}$ (and it is elementary to find examples of unitaries $\hat{U}^{A}$, $\hat{U}^{B}$ such that this is not the case). In the final step of the protocol, the entangled state is measured in appropriate bases, that lead to a violation of a Bell inequality.

The argument presented in Ref.~\cite{Bell4time} is that, given the initial product state of the target systems, local operations that are performed in a definite order would not be able to produce entanglement, even after conditioning on the control system. Therefore, if a set of conditions is satisfied, a violation of Bell inequalities implies that the operations were not performed in a definite order.

Another assumption of the protocol, which ends up a focal point of the present work, is that any additional evolution of the target systems (including their free evolution) between the events of interest can be neglected.

 In the ``Methods'' and the ``Results'' sections, we focus on reproducing the gravitational protocol using special relativistic time dilation. We present an operational setting where the dynamics of all the relevant DoFs is incorporated, in particular the DoFs whose interactions realize the four uniatries $\hat U^A, \hat U^B$. In the ``Ambiguity in the signature of indefinite temporal order'' section we discuss our main result -- that entanglement can be generated and Bell inequality for temporal order can be violated even if temporal order is classically defined -- and argue that it is due to the failure of the additional assumption mentioned above, that target systems have trivial evolution apart from the unitaries marking the four spacetime events of interest. In the ``Conclusion'' section we discuss the implications of our result, including the fundamental question of how to isolate quantum features of a causal structure from other non-classical effects.

\section*{Methods}\label{kinematics} 

\begin{figure}[h!tbp]
\centering
    \includegraphics[width=0.9\linewidth]{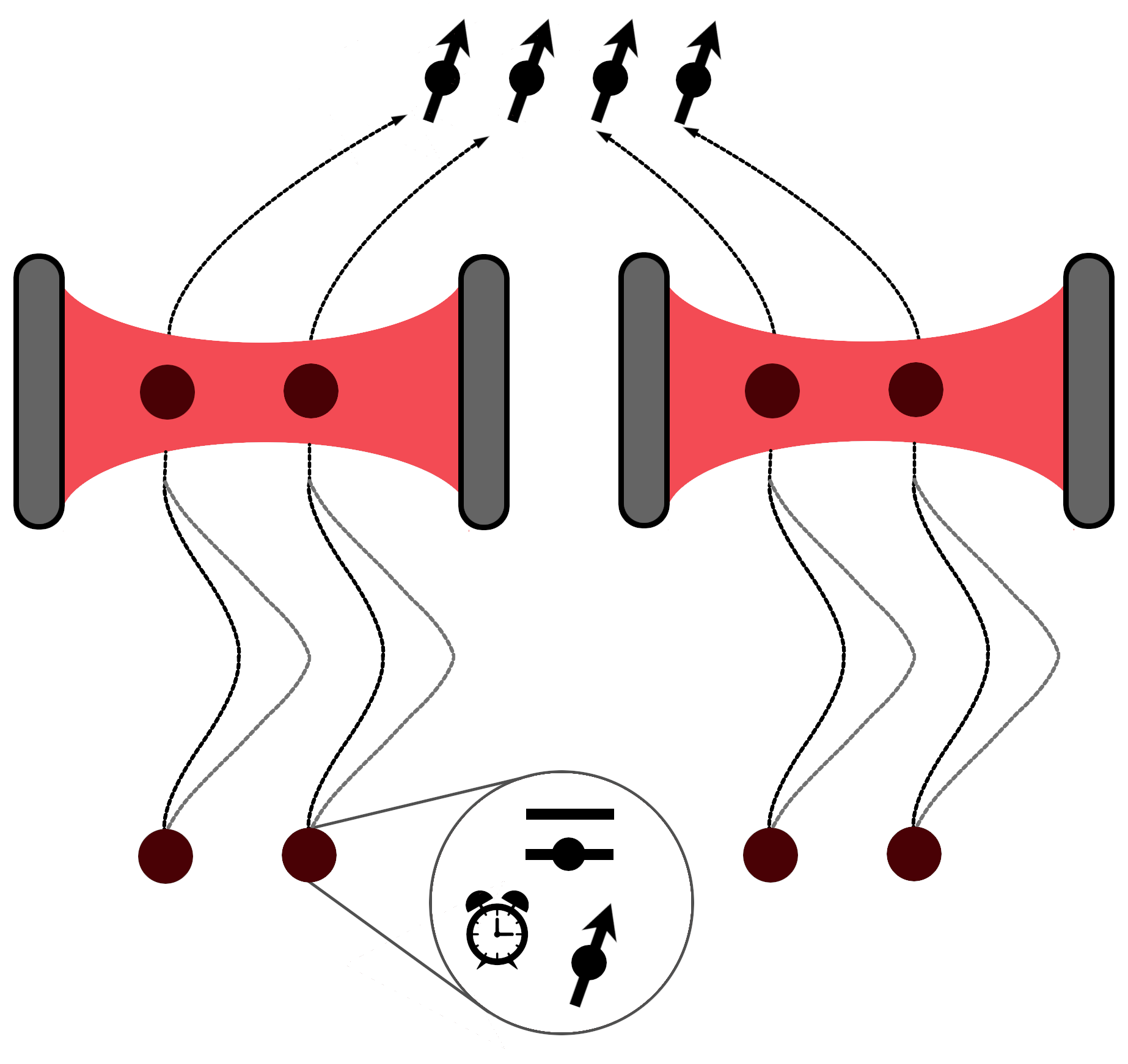}
    \caption{{\bf General scheme of our protocol for a violation of Bell inequalities for temporal order.} In each wing of the experiment we have a quantum field in a cavity, and two composite 'molecules'.  After following entangled pairs of trajectories the molecules interact with the fields at a fixed proper time of their internal clocks. Due to time dilation, entangled state of motion  gets transferred to the order of the interaction events (as well as to other degrees of freedom of the molecules and the cavities).}
    \label{fig:2}
\end{figure}

Instead of using a massive object to control the spacetime geometry---and thus causal order via gravitational time dilation---here we want to control the trajectories of particles so to induce special-relativistic time dilation. The protocol involves several degrees of freedom, which can be best thought of as multiple particles `glued' together (up to the moment when we need to break them apart, as detailed later). We will refer to a bunch of joined particles as a `molecule', although the details of what binds the particles together are irrelevant to the discussion.

Our protocol involves four molecules going through two optical cavities (two molecules per cavity). Each molecule is composed of three particles: a `clock', a `detector', and a `control'. Figure \ref{fig:2} presents a general scheme of our protocol.

The clock is simply a particle with some time-evolving internal state; if the molecule evolves along a classical trajectory, the internal state evolves at a rate proportional to the trajectory's proper time. The role of the clock is to trigger an interaction between the detector and the cavity at the desired proper time. Thanks to the universality of time dilation, the protocol does not depend on the particular mechanism by which the clocks evolve---all we need is that the clock reaches two different orthogonal states depending on the two proper times involved in the protocol  (see, e.g., Refs.~\cite{Zych2011, pikovski2017time}).

The detector is a particle with two internal energy levels that (at the proper time specified by the clock), 
 interacts with a quantum field  confined in a cavity. We use the Unruh-DeWitt detector model for the interaction; see the ``Unruh-DeWitt coupling'' section below for the details of the coupling. 
 
 Finally, the control is a spin-$\frac{1}{2}$ particle, whose two orthogonal spin states serve to define the molecules' trajectories (see, e.g., Ref.~\cite{Margalit2021} for a realisation of coherent spin-dependent trajectories which could be used here). Although each detector interacts with the cavity at the same proper time of its local clock, special-relativistic time dilation implies that the interactions take place at different coordinate times depending on the molecule's trajectory (which, in turn, depends on the spin).

In the protocol, the detectors in each molecule are prepared in their ground state and the clocks are synchronised at a reference starting time, while the spins of different molecules are prepared in an appropriate entangled state (defined below). Each molecule  is sent to a cavity along a trajectory that depends on the spin.  The clock triggers an interaction between the detector in the molecule and the field, creating entanglement between the two. (The proper time at which the interaction happens is chosen such that the molecule is in the cavity for either trajectory). At this point, the molecule is `broken apart': the detector and the clock stay next to the corresponding cavity, while all the controls are brought together in a middle location. A joint measurement on the controls prepares an entangled state of the remaining systems,  which can then be used to violate a Bell inequality.

In our setup, the `target' system on each side, say $S_1$, comprises the field in the cavity as well as the two detectors that go through that cavity. 
Crucially, the field-detector interaction leaves the clock and control unaffected. This is important to ensure that each of the operations  $U^A, U^B$ acts only on the target system. Without this assumption, one could simply entangle each target system with an additional DoF, e.g.~an extra particle, bring the two extra particles together and, by measuring them, induce entanglement on the two target systems. This would be a form of entanglement swapping \cite{Yurke1992} that does not require any control of time ordering (nor any control system for that matter). 

Furthermore, note that since the control DoF goes through the cavity together with the rest of the molecule, we effectively need to trust the involved devices when we assume that the local operations leave the control untouched. This is the reason such a test for indefinite temporal order cannot be formulated in a device-independent way (just as in the gravitational protocol of Ref.~\cite{Bell4time}, see also discussion therein). 

\begin{table*}[th!]
    \begin{tabular}{l|c|c}
            & Gravity   &  Cavity \\ \hline
    Control system & massive body    & spin-$\frac{1}{2}$ particles (one per molecule, two on each side) \\
    Target system & a single 2-level system (e.g., a spin-$\frac{1}{2}$ particle) & Optical cavity mode and two detectors \\
    Local operations & unitaries on each system & Interaction between cavity and detectors
    \end{tabular}
    \caption{\textbf{Comparison between the degrees of freedom involved in the gravitational scheme and ours}. Here we only consider the main scheme from Ref.~\cite{Bell4time} (variations of the scheme involve multiple control systems or different target DoFs.) }
    \label{tab:my_label}
\end{table*}

\begin{figure}
\centering
    \includegraphics[width=0.8\linewidth]{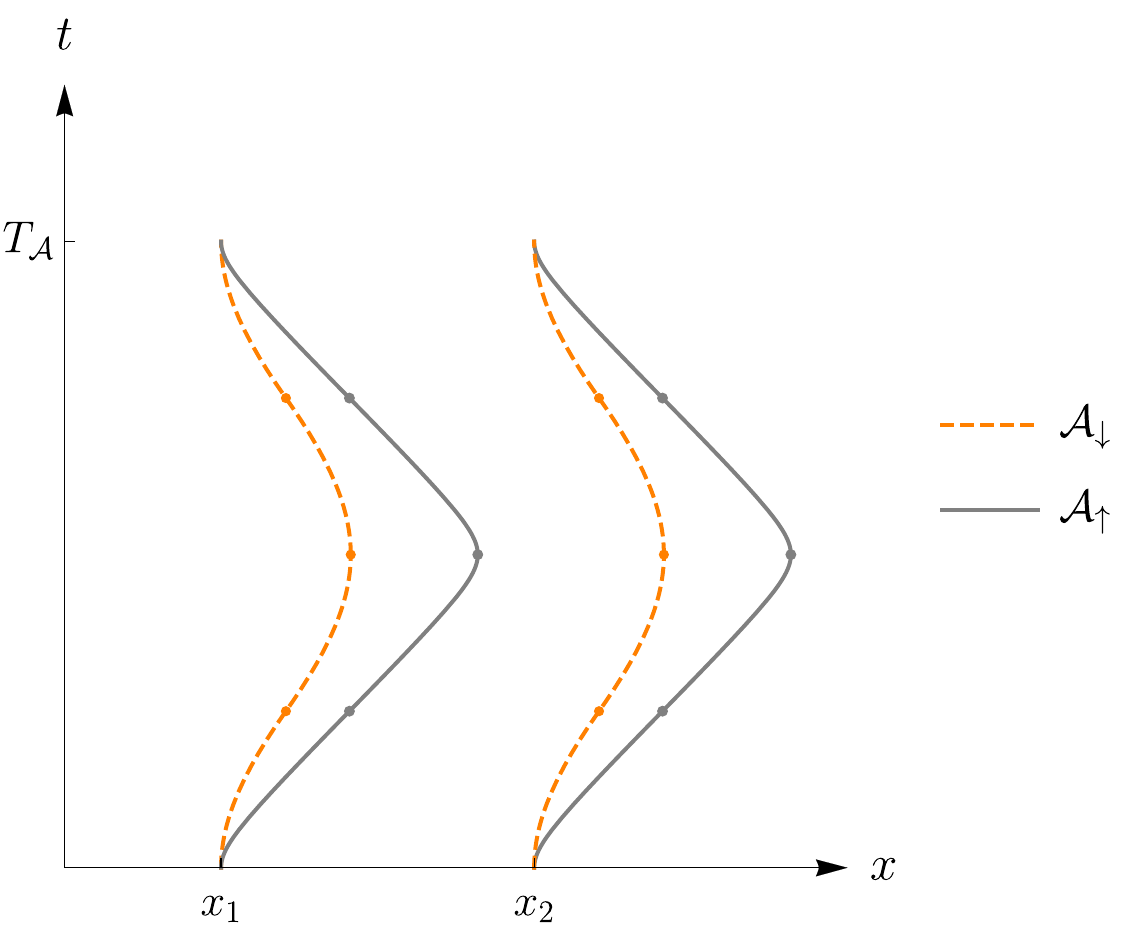}
    \caption{{\bf Trajectories of the molecules in one cavity giving rise to the required time dilation.} $x_1$, $x_2$ are initial positions of the molecules. The trajectory of each molecule depends on the spin of its particle called the `control'. For spin $\ket{\uparrow}$, the trajectory is described by the proper acceleration $\mathcal{A}_{\uparrow}$, and analogously for spin $\ket{\downarrow}$. Dots along the trajectories divide each curve into four geometrically identical hyperbolic segments. The trajectories of the molecules in the second cavity are fully analogous.}
    \label{fig:1}
\end{figure}

\subsection{Trajectories}

Each molecule has two possible trajectories, depending on the spin state. The specific trajectories {we propose} are shown in Fig.~\ref{fig:1}. The involved trajectories are analogous between the two wings of the experiment: They start and end at a common event but the proper time elapsing along each trajectory is different. Both trajectories can be constructed by joining four identical hyperbolic segments, characterised by proper accelerations  $\mathcal{A}_{\uparrow}$,   $\mathcal{A}_{\downarrow}$ for spin up, $\ket{\uparrow}$, or spin down, $\ket{\downarrow}$, respectively.
These segments have to be rotated or flipped according to Fig.~\ref{fig:1}, so that the acceleration for each trajectory switches sign three times. Let us assume that the acceleration for the spin $\ket{\uparrow}$ is bigger than an acceleration for the spin $\ket{\downarrow}$. Knowing the value of proper time along the generic hyperbolic trajectory (see e.g.~Ref.~\cite{Dragan2021}), we can find the difference between proper times measured at the common end of one such pair of trajectories: 
\begin{gather}
    \Delta\tau=
    \underbrace{
    \frac{4}{\mathcal{A}_{\downarrow}}\mathrm{asinh}(\frac{\mathcal{A}_{\downarrow} T_{\mathcal{A}}}{4})
    }_{\tau_{\downarrow}}
    -
    \underbrace{
    \frac{4}{\mathcal{A}_{\uparrow}}\mathrm{asinh}(\frac{\mathcal{A}_{\uparrow} T_{\mathcal{A}}}{4})
    }_{\tau_\uparrow},
\end{gather}
where $\tau_{\uparrow/\downarrow}$ is the proper time measured along trajectory of a $\uparrow/\downarrow$ spin and $T_{\mathcal{A}}$ is the total travel time in a common inertial frame of reference, e.g.~the frame used to initially synchronize the clocks.

Note that it is possible to achieve a large value of $\Delta\tau$ using a small value of  proper acceleration.
For a large value of $T_{\mathcal{A}}$ we have:
\begin{gather}
\Delta\tau=
2\left(
\frac{\log(\frac{\mathcal{A}_{\downarrow}^2}{4})}{\mathcal{A}_{\downarrow}}
-
\frac{\log(\frac{\mathcal{A}_{\uparrow}^2}{4})}{\mathcal{A}_{\uparrow}}
\right)\\
+
4\left(
\frac{1}{\mathcal{A}_\downarrow}
-
\frac{1}{\mathcal{A}_\uparrow}
\right)\log(T_{\mathcal{A}})+\mathcal{O}\left(\frac{1}{T_\mathcal{A}^2}\right).
\end{gather}
It means that we can achieve the required $\Delta\tau$ (to make the events defined by such time dilated pair of clocks timelike) by accelerating for long enough time, even for arbitrarily small accelerations.
Finally, taking the initial state of the spin of the two molecules to be entangled, one of the two clocks in each cavity will  ``be older'' than another in a correspondingly correlated manner. Thus, the order of operations, controlled by the clocks, will likewise be ``entagled'', as a direct consequence of the initial spin entanglement and of time dilation. In this scenario, the joint spin state of the two molecules plays the role of the control, which is played by the position of a massive object in the gravitational case~\cite{Bell4time}. Table \ref{tab:my_label} summarises the differences between the gravitational and our cavity-based implementation of the protocol, in terms of control systems, target(s), and local operations.

\subsection{Unruh-DeWitt coupling}\label{sec:UdWcoupling}

\subsubsection{Interaction between a detector and a cavity}
\label{UDWcoupling}
In this section we define the interaction between a two-level system, the `detector', and the scalar field inside the cavity via the pointlike Unruh-DeWitt Hamiltonian \cite{Unruh1976,Unruh1984}. We first discuss key properties of the field operators.

We consider a scalar field of a mass $m$ governed by the Klein-Gordon equation\footnote{For generality, we write everything for arbitrary $m$, although in the numerical calculations below we set $m=0$.},
\begin{equation}
 \left(\Box+m^2  \right) \phi=0,
 \end{equation}
in a cavity of length $L$ fulfilling Dirichlet boundary conditions, $\phi(x=0)=\phi(x=L)=0$. The field has the following mode solutions:
\begin{equation}
u_n(x,t)=\frac{1}{\sqrt{\omega_n L}} \sin{\left(k_n x\right)}e^{-i\omega_n t}\equiv u_n(x)e^{-i\omega_n t},    
\end{equation}
where $\omega_n=\sqrt{k_n^2+m^2}$, $k_n=\frac{n\pi}{L}$, $n \in \mathbb{N}$.
Using these modes, the field operator $\hat{\phi}$ can be decomposed as:
\begin{equation}
\hat{\phi}(x)=\sum_{n}\left[\hat{a}_n^{\dagger} u_{n}(x)+\hat{a}_n u_n(x)\right],
\end{equation}
where $\hat{a}_n$ and $\hat{a}_n^{\dagger}$ are annihilation and creation bosonic operators satisfying the canonical commutation relations, $\left[\hat{a}_n,\hat{a}_k^{\dagger}\right]=\delta_{nk}$ and $\left[\hat{a}_n,\hat{a}_k\right]=\left[\hat{a}_n^{\dagger},\hat{a}_k^{\dagger}\right]=0$.

For the detector we consider a two-level system, the simplest model of an atom,  with an energy gap $\Omega$, and position parameter denoted $x_d$, (where the subscript $d$ hereafter stands for the 'detector'). The full Hamiltonian consists of the free Hamiltonians of the scalar field and the detector, and an interaction Hamiltonian. 
One of the simplest  choices of the interaction between a scalar field and a two-level system is the pointlike Unruh-DeWitt (UDW) Hamiltonian which in the Schr\"odinger picture has the following form:
\begin{equation}
\hat{H}_{\mathrm{UDW}}=\lambda~\chi_{d}(t)~\hat{\mu}_{\mathrm{S}}~\hat{\phi}(x_d),
\label{eqn:UdWhamiltonian}
\end{equation}
where $\lambda$ is a dimensionless coupling constant; the real function $\chi_{d}(t)$ is equal to $0$ when the detector does not interact and $1$ for any other time and is commonly referred to as the switching function; $\hat{\mu}_{\mathrm{S}}$ is the monopole operator $\hat{\mu}_{\mathrm{S}}=\hat{\sigma}^{+}+\hat{\sigma}^{-}=\ket{g}\bra{e}+\ket{e}\bra{g}$, where $\ket{g}$ is the ground state of the two-level system and $\ket{e}$ is its excited state. The Hilbert space spanned by $\ket{g}$ and $\ket{e}$ will be called the internal Hilbert space of the detector. Finally, $\hat{\phi}(x_d)$ is the field operator evaluated at the position of  the detector. 

As mentioned above, the full Hamiltonian includes also the time-independent free Hamiltonian of the  field and of the two-level system, which reads $\hat{H}_0=\sum_n \omega_n \hat{a}_{n}^{\dagger}\hat{a}_{n} \otimes
\mathds{1}+\mathds{1} \otimes \Omega \hat{\sigma}^{+} \hat{\sigma}^{-}$. 
Thus, the evolution of the state of the full system in the Dirac picture (also called the interaction picture) is here given by the unitary of the form: 
\begin{equation}
\hat{U}
=\mathcal{T}\exp{
-i\int_{-\infty}^{\infty} 
\mathrm{d}t\hat{H}_{\mathrm{UDW}}^{(D)}(t)
},
\label{eqn:U}
\end{equation}
where $(D)$ stands for the Dirac picture and $\mathcal{T}$ is the time-ordering operator. It can be shown that \cite{RevModPhys.80.787}: 
\begin{align}
\hat{H}_{\mathrm{UDW}}^{(D)}(t)=\chi(t)~\lambda~\hat{\mu}^{(D)}~\hat{\phi}^{(D)},
\end{align}
where:
\begin{eqnarray}
\hat{\mu}^{(D)}&=&\left(e^{i\Omega t} \hat{\sigma}^{+}+e^{-i\Omega t} \hat{\sigma}^{-}\right),\\ \label{intpicfield}
\hat{\phi}^{(D)}(x_d)&=&\sum_{n}\left(\hat{a}_n^{\dagger} u_{n}(x_d)e^{i\omega_n t}+H.c.\right).
\end{eqnarray}
The evolution operator \eqref{eqn:U} can be expanded into the Dyson series. For small value of the coupling constant $\lambda$ we  limit this series only to the first order, the second order expansion does not change our results as shown in Supplementary Note 1. Thus, in the above approximation:
\begin{gather}
\hat{U}
=
\mathds{1}-i
\lambda\int_{-\infty}^{\infty} 
\mathrm{d} t
~\chi_{d}(t)
\left(e^{i\Omega t} \hat{\sigma}^{+}+e^{-i\Omega t} \hat{\sigma}^{-}\right)
\nonumber
\\
\times\sum_{n}\left(\hat{a}_n^{\dagger} u_{n}(x_d)e^{i\omega_n t}+
\hat{a}_n u_{n}(x_d)e^{-i\omega_n t}\right).
\label{eqn:operatorU}
\end{gather} 

\subsubsection{Two molecules in the cavity}\label{onecavity}

Before moving to the full protocol, let us consider the interaction between one cavity and two molecules. Each molecule contains a two-level detector, which interacts with the cavity via the UDW Hamiltonian \eqref{eqn:UdWhamiltonian}, plus a clock and a control spin. At the beginning of the protocol, the cavity field is in the vacuum, denoted $\ket{0}$, and the two detectors are in their respective ground states. We synchronize the clocks, namely we prepare them in the same state $\ket{\tau_0}$, and we prepare the two control spins in the entangled state $\frac{1}{\sqrt{2}}(\ket{\updownarrows}+\ket{\downuparrows})$. After the spin-dependent accelerations described in the ``Trajectories'' section, the joint state of cavity and molecules is

\begin{gather}\label{twomoleculestate}
\frac{1}{\sqrt{2}}(\ket{\tau_{\uparrow}\tau_{\downarrow}}\ket{\updownarrows}+\ket{\tau_{\downarrow}\tau_{\uparrow}}\ket{\downuparrows})\ket{gg}\ket{0},
\end{gather}
where $\ket{\tau_{\uparrow/\downarrow}}$ is the state of the clock after it evolved for time $\tau_{\uparrow/\downarrow}$.
Note that for later convenience we have grouped the degrees of freedom by type (clock, spin, detector), rather than by molecule.  For example, instead of
\begin{align}
\underbrace{\ket{\tau_{\uparrow}}\ket{\uparrow}\ket{g}}_{1^{st}~molecule}\otimes\underbrace{\ket{\tau_{\downarrow}}\ket{\downarrow}\ket{g}}_{2^{nd}~molecule},
\end{align}
we write $\ket{\tau_{\uparrow}\tau_{\downarrow}}\ket{\updownarrows}\ket{gg}$, where inside each `ket' symbol we write first the state relative to molecule 1 and then the state relative to molecule 2.

We see from Eq.~\eqref{twomoleculestate} that, after the spin-dependent accelerations, the clocks become entangled with the remaining systems. This can be problematic because, in the final protocol, we aim to observe entanglement in the target systems (detectors+cavities) after projecting the control in an appropriate state. Anything that correlates to the target, including the clocks, would effectively degrade entanglement.
We can circumvent this by re-synchronising the clocks after they had passed through the cavity. This can be achieved, for example, by flipping the molecules' spins and imparting the accelerations identical to those in the first phase, whereby at the end, all the trajectories accrue equal proper times (this is analogous to the decorrelation of clocks in the gravitational scenario, discussed in Ref.~\cite{Bell4time}). {This procedure allows us to ignore the clocks in the final state. Therefore, we simplify the notation and map} $\ket{\tau_{\uparrow/\downarrow}}\ket{\uparrow/\downarrow}\mapsto \ket{\uparrow/\downarrow}$ which allows us to replace state \eqref{twomoleculestate} with
\begin{align}
    \ket{\Psi_0^{\mathrm{1}}}=
    \frac{1}{\sqrt{2}}(\ket{\updownarrows}+\ket{\downuparrows})\ket{gg}\ket{0},
\end{align}
where the subscript $0$ informs that this is the initial state of the system (before detectors and cavity interact) and the index $1$ in the superscript informs us that this is the state of the field and two molecules in cavity number $1$ (see Fig.~\ref{fig:2}).

Note that according to the scheme shown in the ``Interaction between a detector and a cavity'' section, the state $\ket{\updownarrows}$ corresponds to the case when the detector at position $x_2$ is the first to interact with the cavity (the spin of the molecule that interacts earlier is $\downarrow$). Similarly, for state $\ket{\downuparrows}$, the detector at $x_1$ interacts earlier than detector at $x_2$. In further calculations, we assume that the duration $T$ of each detector's interaction with the cavity is smaller than the time dilation between the two  trajectories, i.e.,  $T\leq\Delta\tau$, to ensure that the two detectors do not interact simultaneously from the perspective of the cavity's reference frame.

As the interaction between each detector and the cavity is described by Eq.~\eqref{eqn:U} and knowing that the order of interactions is given by the molecule's spin, we can write the final state in the form:
\begin{gather}\label{cavityswitch}
\ket{\Psi^{\mathrm{1}}}
=
\frac{1}{\sqrt{2}}\left(\ket{\updownarrows}\hat{U}_1\hat{U}_2\ket{g}\ket{g}\ket{0}+\ket{\downuparrows}\hat{U}_2\hat{U}_1\ket{g}\ket{g}\ket{0}\right).
\end{gather}
Here, operators $\hat{U}_1$ and $\hat{U}_2$ act on the internal states of the first (left) and the second (right) detector respectively. For ease of notation we define the state $\ket{\psi_R}:=\hat{U}_1\hat{U}_2\ket{g}\ket{g}\ket{0}$, where the subscript $R$ denotes that the right detector interacts before the left one, and analogously define $\ket{\psi_L}:=\hat{U}_2\hat{U}_1\ket{g}\ket{g}\ket{0}$. We thus  have:
\begin{gather}\label{onecavityafterinteraction}
\ket{\Psi^{\mathrm{1}}}
=
\frac{1}{\sqrt{2}}\ket{\updownarrows}\ket{\psi_R}+\frac{1}{\sqrt{2}}\ket{\downuparrows}\ket{\psi_L}.
\end{gather}
Using the evolution operator \eqref{eqn:U} we find an explicit form of $\ket{\psi_R}$, $\ket{\psi_L}$, up to leading order in the interaction parameter $\lambda$ using Eq.~\eqref{eqn:operatorU} (see Supplementary Note 2 for details).
The results of this calculation are 
\begin{widetext}
\begin{gather}
\ket{\psi_R}=
\ket{gg}\ket{0}+\ket{ge}\ket{\phi_{ge}^R}+\ket{eg}\ket{\phi_{eg}^R}+\mathcal{O}(\lambda^2),
\label{eqn:psiR}
\end{gather}
\begin{gather}
\ket{\psi_L}=
\ket{gg}\ket{0}+\ket{ge}\ket{\phi_{ge}^L}+\ket{eg}\ket{\phi_{eg}^L}+\mathcal{O}(\lambda^2),
\label{eqn:psiL}
\end{gather}
\end{widetext}
where $\ket{\phi_{ge}^{L/R}}$ and $\ket{\phi_{eg}^{L/R}}$ are first order in $\lambda$, describing field states containing a single excitation, see Supplementary Note 2 for their explicit expression. 

\section{Results} \label{results}
\subsection{Protocol -- two cavities, four molecules}\label{twocavities}
We now move to the full protocol, whose aim is to demonstrate indefinite causal order in a fully operational setting.
We consider two `wings' of the experiment, where in each wing two operations take place. The goal is to correlate the order of each pair of operations with a control system and, after measuring the control, produce an entangled state between the two wings, which would not have been possible had the operations been realised in a definite order.

Recall that we want to entangle the `target' systems which comprise the fields in both cavities as well as the detectors (which are two per cavity), while the four spin-$\frac{1}{2}$ particles play the role of the control. The full protocol involves  four molecules (two in each wing), where each molecule contains a clock, a detector, and a spin-$\frac{1}{2}$ particle. Initially, the state of these molecules reads:
\begin{align}
\ket{\tau_0\tau_0\tau_0\tau_0}\frac{1}{\sqrt{2}}(\ket{\updownarrows\updownarrows}+\ket{\downuparrows\downuparrows})\ket{gggg},
\end{align}
where $\ket{\tau_0}$ is the initial state of one clock (and thus all four clocks are initially decorrelated and synchronised) and each detector is in its ground state, while the spins are entangled.

Note that this state is constructed in such a way that the four molecules can be divided into two identical pairs. These pairs are then accelerated as explained in the ``Trajectories'' section. 
After the process of acceleration, the state of all four molecules is:
\begin{align}
\frac{1}{\sqrt{2}}(\ket{\tau_\uparrow\tau_\downarrow\tau_\uparrow\tau_\downarrow}\ket{\updownarrows\updownarrows}+\ket{\tau_\downarrow\tau_\uparrow\tau_\downarrow\tau_\uparrow}\ket{\downuparrows\downuparrows})\ket{gggg}.
\end{align}
Due to time dilation, after the acceleration takes place, one clock from each pair of molecules will be older than the other from the same pair. We cannot determine which one because the initial state of spins is a superposition of two different possibilities. Next, each pair of molecules enters a cavity. The first pair is placed in cavity $1$ at the respective positions $x_1$ and $x_2$ (determined relative to the boundary of this cavity at $x=0$). The second pair of molecules is placed in cavity $2$ at the same positions  relative to that cavity. Recall that the field in each cavity is initially in the vacuum state. As discussed in the ``Two molecules in the cavity'' section, we can decorrelate the  clock DoF by re-synchronising the clocks. This is done after the molecules interact with the cavities. To keep track of the effect of the clocks on the detector-field interactions, it is enough to note that a detector belonging to a molecule with spin $\downarrow$ interacts first (and with spin $\uparrow$ interacts second). To sum up the protocol, the initial state of the total system including the two cavities is:
\begin{gather}
    \ket{\Psi_{0}^{\mathrm{tot}}}=\frac{1}{\sqrt{2}}(\ket{\updownarrows\updownarrows}+\ket{\downuparrows\downuparrows})\underbrace{\ket{gg}\ket{0}}_{\in S_1}\otimes\underbrace{\ket{gg}\ket{0}}_{\in S_2},
\end{gather}
where we used $\otimes$ to separate states from the two cavities and showed which degrees of freedom comprise the targets $S_1$ and $S_2$ introduced in the ``Methods'' Section.

After the interactions between the detectors and the cavity fields, the state of the system reads:
\begin{gather}
    \ket{\Psi^{\mathrm{tot}}}
    =
    \frac{1}{\sqrt{2}}\left(\ket{\updownarrows\updownarrows}\ket{\psi_R}\otimes\ket{\psi_R}+
    \ket{\downuparrows\downuparrows}\ket{\psi_L}\otimes\ket{\psi_L}\right),
\end{gather}
where the detectors and cavity states, $\ket{\psi_{L/R}}$, are those in  Eqs~\eqref{eqn:psiR}, \eqref{eqn:psiL}.

Next the molecules are `broken apart'. All  spins are send to a common location where, at an event labelled $D$, they are jointly measured in the basis  $\frac{1}{\sqrt{2}}(\ket{\updownarrows\updownarrows}\pm\ket{\downuparrows\downuparrows})$.  This prepares the remaining systems--- cavities and detectors (which stay next to their cavities)---in the state
\begin{align}
\ket{\Psi^{\pm}}
=
\frac{1}{\sqrt{2}}\left(\ket{\psi_R}\otimes\ket{\psi_R}\pm\ket{\psi_L}\otimes\ket{\psi_L}\right),
\end{align}
where the sign $\pm$ depends on the measurement outcome. 
This state is entangled as long as $\ket{\psi_L}\neq\ket{\psi_R}$.
This is typically shown by finding the scalar product between $\ket{\psi_L}$ and $\ket{\psi_R}$ which requires the Dyson series up to second order in $\lambda$. 

However, there is an equivalent but technically much simpler method to show that the final state is in general entangled,
which is to consider another measurement performed on each detector pair in a basis containing the vector $1/\sqrt{2}\left(\ket{ge}+\ket{eg}\right)$.  
The resulting conditional states of the fields (arising from Eqs~\eqref{eqn:psiR} and \eqref{eqn:psiL}) read
\begin{gather}
\frac{1}{\sqrt{2}}(\bra{ge}+\bra{eg})\ket{\psi_{R}}
=
\frac{1}{\sqrt{2}}\left(\ket{\phi_{ge}^{R}}+\ket{\phi_{eg}^{R}}\right)+\mathcal{O}(\lambda^2),
\\
\frac{1}{\sqrt{2}}(\bra{ge}+\bra{eg})\ket{\psi_{L}}
=
\frac{1}{\sqrt{2}}\left(\ket{\phi_{ge}^{L}}+\ket{\phi_{eg}^{L}}\right)+\mathcal{O}(\lambda^2).
\end{gather}
Upon measuring  the control in the above defined entangled basis, $\frac{1}{\sqrt{2}}(\ket{\updownarrows\updownarrows}\pm\ket{\downuparrows\downuparrows})$, the joint state of two cavities fields up to the leading order is:
\begin{widetext}
\begin{align}
\ket{\Tilde{\Psi}^{\pm}}\propto
\underbrace{
\left(\ket{\phi_{ge}^{R}}+\ket{\phi_{eg}^{R}}\right)
}_{\ket{\Phi_{R}}}\otimes
\underbrace{
\left(\ket{\phi_{ge}^{R}}+\ket{\phi_{eg}^{R}}\right)
}_{\ket{\Phi_{R}}}
\pm
\underbrace{
\left(\ket{\phi_{ge}^{L}}+\ket{\phi_{eg}^{L}}\right)
}_{\ket{\Phi_{L}}}
\otimes
\underbrace{
\left(\ket{\phi_{ge}^{L}}+\ket{\phi_{eg}^{L}}\right)
}_{\ket{\Phi_{L}}}.
\label{eqn:jointstate}
\end{align}
\end{widetext}

Note that the states $\ket{\Phi_L}$ and $\ket{\Phi_R}$ describe only the fields in the cavities. The conditional state $\ket{\Tilde{\Psi}^\pm}$ is entangled as long as $\ket{\Phi_R}\neq\ket{\Phi_L}$ which is easier to verify than the condition $\ket{\psi_R}\neq\ket{\psi_L}$. We further note that the measurement on the detectors has different possible outcomes, not all of which result in entangled field states. 
Crucially however, for a product state across $S_1$ and $S_2$, all local detector measurement would result in product states of the fields. Therefore, obtaining entangled state for one of the  outcomes is sufficient to prove that the original state was entangled.  

To sum up, we have shown that using two cavities, four molecules, and the basic effects of special relativity, we can obtain the state given in Eq.~\eqref{eqn:jointstate}. If this state is entangled,  in the next steps of the protocol this entanglement can be used to violate Bell's inequalities with the goal to prove that the detector--field interaction events were not classically  ordered.

\subsection{Entanglement of the final state}\label{sec:surprising_result}
\begin{figure}[h]
\centering
    \includegraphics[width=1\linewidth]{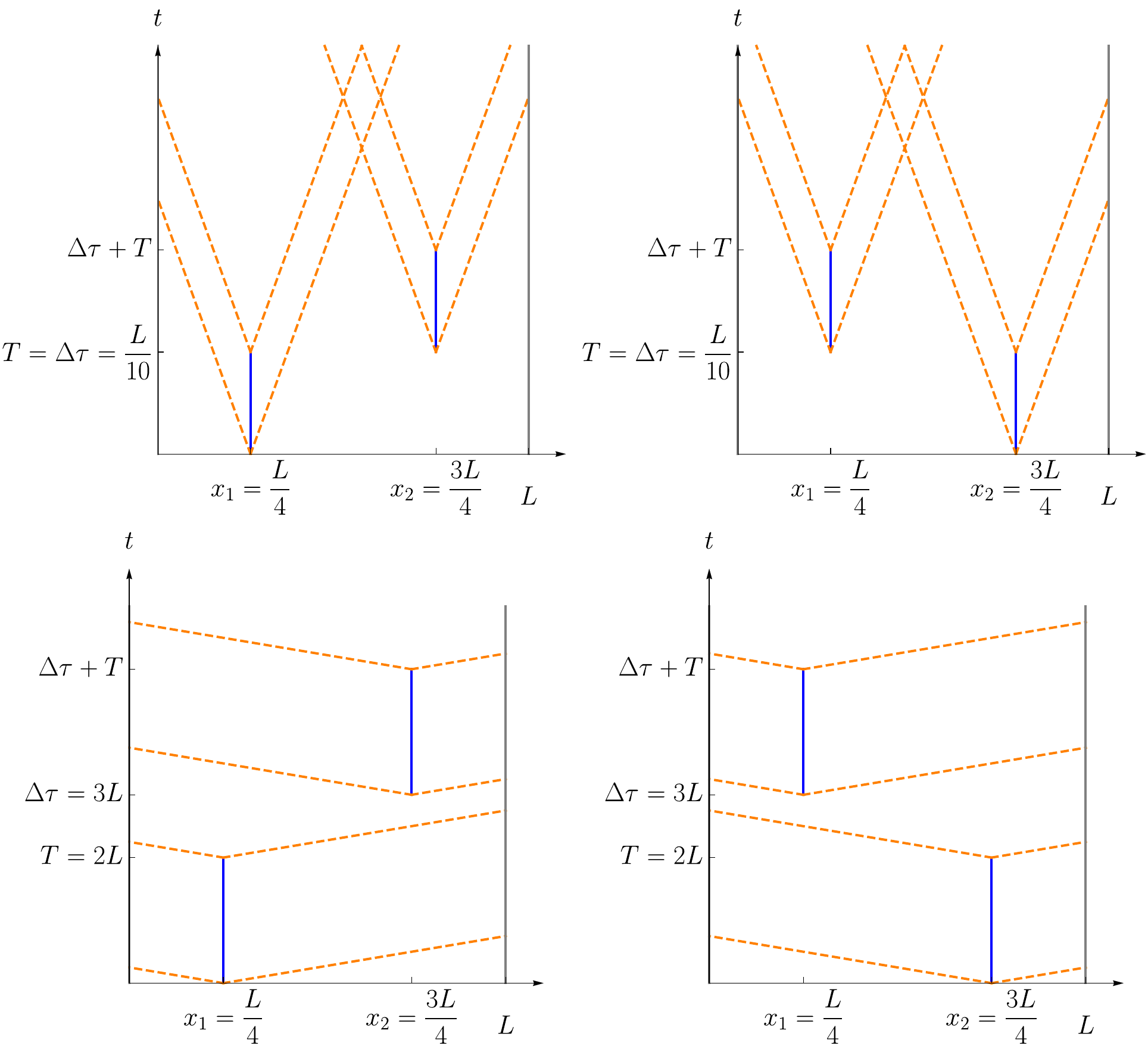}
        \caption{{\bf Spacetime diagrams of interacting detectors.} Spacetime diagram showing the regions where the detector-field interactions lead to an entangled state, Eq.~\eqref{eqn:jointstate}. The diagrams are for one cavity --  the interaction regions are idenatically defined for the second cavity. The first row shows  spacelike separated regions which nevertheless yield entanglement. The second row shows timelike separation which yields maximal entanglement. The columns correspond to the two amplitudes of the process that are superposed using the control (spin) state. See Supplementary Note 3 for the supporting calculations.}
        \label{fig:grid}
\end{figure}

We now proceed to prove  that the state in Eq.~\eqref{eqn:jointstate} can indeed be entangled by showing that there exist parameters for which $\ket{\Phi_R}\neq\ket{\Phi_L}$. The scalar product between these states reads:
\begin{widetext}
\begin{align}
\braket{\Phi_R}{\Phi_L}
=
\frac{
    \sum_{k}
    \frac{e^{-i\Delta\tau(\omega_k+\Omega)}}{(\omega_k+\Omega)^2}
    \left[
    1-\cos\left(T(\omega_k+\Omega)\right)
    \right]
    \left(
    u_k(x_1)
    +
    e^{i\Delta\tau(\omega_k+\Omega)}
    u_k(x_2)
    \right)^2
}{
2
    \sum_{n}
    \frac{\sin^2\frac{T(\omega_n+\Omega)}{2}}{(\omega_n+\Omega)^2}
    \left[
    u_n(x_2)^2
    +
    2u_n(x_2)u_n(x_1)\cos\left(\Delta\tau(\omega_n+\Omega)\right)
    +u_n(x_1)^2
    \right]
}\,.
\end{align}
\end{widetext}
Explicit derivation is presented in Supplementary Note 3. Therein we further show how to choose parameters for which the states are not just different but orthogonal. To summarise this procedure, the parameters that have to be chosen are:
\begin{enumerate}
    \item The length $L$ of each cavity and the positions $x_1, x_2$ of the molecules relative to their respective cavity.
    \item The energy gap $\Omega$ of the detectors.
    \item The duration $T$ of the interaction between each detector and the cavity field.
    \item The time dilation $\Delta \tau$ between clocks within each pair of molecules (within each cavity), arising from the different accelerations $ \mathcal{A}_{\uparrow/\downarrow}$.
\end{enumerate}
{Remarkably, it is also possible to find parameters such that $\ket{\Phi_R}\neq\ket{\Phi_L}$ even when the two field-detector interactions within each cavity are space-like separated, see Fig.~\ref{fig:grid}. This means that it is possible to obtain an entangled state also in the  case when, in some reference frame, the relevant operations are performed in the same temporal order. Although from the perspective of such a reference frame the operations would take place over four different coordinate regions, their order would be the same in each of the amplitudes. Obviously then,  entanglement generated in this scheme cannot be simply attributed to non-classical temporal order}.  We discuss  implications and argue for generality of this result in the next sections.

\subsection{Ambiguity in the signature of indefinite temporal order
} \label{Bell4time}

We have explicitly modelled a special-relativistic version of a protocol where gravitational time dilation and quantum superposition lead to an indefinite temporal order of events from Ref.~\cite{Bell4time}. This protocol was formulated in terms of a `Bell inequality for temporal order'. The idea was to formulate a protocol where operations in a \textit{definite} order cannot produce an entangled state, if an appropriate set of assumptions is satisfied. (The final step of the protocol requires measurements on the state in order to violate a Bell inequality and verify the entanglement.) The motivation was to find a test of temporal order that is  \textit{theory independent}, which was based on the observation that a violation of a Bell inequality would prove indefinite temporal order without assuming that the final state is described by quantum mechanics. 
In this section, we reexamine the assumptions made in Ref.~\cite{Bell4time}, showing that the theory-independent nature of the argument is problematic.

We first remark that the issue we have identified does not arise if we believe that quantum mechanics is valid, i.e., if we wish to provide experimental evidence for non-classical temporal order assuming that the involved states and transformations are faithfully described by  quantum theory. In such a case, already the one-cavity part of our protocol, the ``Methods'' section, or the gravitational equivalent, is sufficient. The reason is that at an abstract level, these protocols implement a ``quantum switch'' \cite{chiribella09} -- a scenario where   two local operations, represented by unitary operators $\hat{U}^A$ $\hat{U}^B$, act on a target system in an order determined by a control system which is prepared in a superposition, thus producing a final state of the form
\begin{multline}\label{switchstate}
\ket{\psi_{\textrm{fin}}}= \\ \frac{1}{\sqrt{2}}\left(\ket{0}\hat{U}^A\hat{V}_0\hat{U}^B+\ket{1}\hat{U}^B\hat{V}_1\hat{U}^A\right)
\ket{\psi}.
\end{multline}
Here,  $\ket{0}$, $\ket{1}$ are two basis states of the control, and $\ket{\psi}$ is the initial state of the target system.  $\hat{V}_0$ and $\hat{V}_1$ are two arbitrary unitary operators, representing the evolution of the target between the two operations. Most presentations of the switch do not include the intermediate evolution, but we will see shortly that this is important in our context\footnote{For generality, one can also include a control-dependent initial state, but this is not necessary for our analysis.}. Given an implementation of the switch, by making appropriate final measurements for a set of suitably chosen operations $\hat{U}^A$, $\hat{U}^B$ it is possible to prove that the operations are not performed in a definite order. This procedure is known as measuring a \textit{causal witness} \cite{araujo15, Branciard2016a} and it could in principle be incorporated within a single-cavity variant of our protocol to demonstrate that, if the quantum description  of the experiment is correct, the cavity-detector interactions do not take place in a definite order.

Returning to the Bell inequality approach, it essentially is an `entangled' version of the switch, with the final state of the form
\begin{multline}\label{entangledswitch}
    \ket{\psi_{\textrm{fin}}} =\\ \frac{1}{\sqrt{2}}\left(\ket{0}\hat{U}^{A_1}\hat{V}_0\hat{U}^{B_1}\ket{\psi}\otimes
    \hat{U}^{A_2}\hat{V}_0\hat{U}^{B_2}\ket{\psi} \right.\\
    +
    \left.
    \ket{1}\hat{U}^{B_1}\hat{V}_1\hat{U}^{A_1}\ket{\psi} \otimes
    \hat{U}^{B_2}\hat{V}_1\hat{U}^{A_2}\ket{\psi}\right).
\end{multline}

In order to use such a state to disprove classical temporal order among $\hat{U}^A$, $\hat{U}^B$, the state has to arise in a scenario satisfying  all assumptions used to derive the Bell inequality for temporal order,  apart from the assumption of classical order itself. All assumptions used in Ref~\cite{Bell4time} for the derivation of the inequality are: the initial state of the target systems $S_1,S_2$ is separable; transformations performed on the targets are local (i.e.~the operations on target $S_j$  act as identity on other degrees of freedom of the system); events/spacetime regions at which transformations and measurements take place are suitably separated: both interaction events in one wing are spacelike separated from both interaction events in the other wing, and event $\mathrm D$ is spacelike from the events at which the Bell measurements are performed (which as, usual in Bell inequalities, are assumed to be spacelike from each other); the choices of bases for Bell measurements are independent of all other aspects of the experiment (often referred to as `free choice' assumption); and, finally, of course the assumption that events at which transformations and measurements are performed are classically ordered.

The surprising result identified in the ``Entanglement of the final state'' section is that entanglement is produced while \textit{all} the above assumptions are met, including the assumption of classical order. Clearly, some other assumption was made to derive the inequality and is violated in our implementation. Indeed, as already mentioned, the additional assumption  made in Ref.~\cite{Bell4time} is that the target systems do not have non-trivial evolution apart from the transformations $\hat U^A$ and $\hat U^B$. Below we explain why this assumption is violated in the present implementation and in the ``Ambiguity in the signature of indefinite temporal
order'' section we argue that this  will remain true in a generic dynamical implementation.

In order to see  where the assumption of no free evolution enters and why it is the culprit it is again sufficient to look at one wing of our setup. Comparing a single-cavity scenario, Eq.~\eqref{cavityswitch}, and a generic quantum switch, Eq.~\eqref{switchstate}, one finds that the evolution operators $\hat{U}_1$, $\hat{U}_2$ in Eq.~\eqref{cavityswitch} are not directly representing the local operations $\hat{U}^A$, $\hat{U}^B$. The reason is that $\hat{U}_1$, $\hat{U}_2$ are written in the Dirac (interaction) picture, which necessarily includes time evolution with respect to the free Hamiltonian (starting from some initial time established in a common reference frame). By unravelling this time evolution, one finds precisely a state of the form \eqref{switchstate}, where $\hat{V}_0$ and $\hat{V}_1$ represent the free evolution of the targets (i.e.~cavity and detectors) between the interactions. Note that the time intervals between the events, and thus intervals of free evolution,  are equal in the reference frame of the cavity, which means that in that frame $\hat V_0=\hat V_1$. On the other hand, $\hat{U}^A$, $\hat{U}^B$ describe \textit{only} the field-cavity interactions in the Schr{\"o}dinger picture\footnote{Strictly speaking, we should add free-evolution operators also before and after the two local unitaries---not only in between. However, free evolution acts trivially on the vacuum state (our initial state), while the final evolution can be reabsorbed in the definition of the measurement basis.}.

The reason this is relevant here is that if the free Hamiltonian 
does not commute with the interaction, the free evolution does not commute with $\hat U^A, \hat U^B$, and this is how the presence of entanglement can be explained  in the state in Eq.~\eqref{entangledswitch} in a frame where $\hat V_0=\hat V_1$ and the events are spacelike separated. Note that in that case there is a reference frame where  $A_j\prec B_j$ for both states of the control, however, as we mentioned above, in that frame necessarily $\hat V_0 \neq \hat V_1$, since the time intervals of free evolution along the worldlines of the molecule 1 and 2 are in such a frame necessarily different. In such a frame the final state in the two wing scenario becomes 
\begin{multline}\label{finalordered}
    \ket{\psi_{\textrm{fin}}} =\\ \frac{1}{\sqrt{2}}\left(\ket{0}\hat{U}^{B_1}\hat{V}_0\hat{U}^{A_1}\ket{\psi}\otimes
    \hat{U}^{B_2}\hat{V}_0\hat{U}^{A_2}\ket{\psi} \right.\\
    +
    \left.
    \ket{1}\hat{U}^{B_1}\hat{V}_1\hat{U}^{A_1}\ket{\psi} \otimes
    \hat{U}^{B_2}\hat{V}_1\hat{U}^{A_2}\ket{\psi}\right),
\end{multline}
and the presence of entanglement is thus interpreted as due to overall  different dynamics depending on the control, $\hat{U}^{B_i}\hat{V}_0\hat{U}^{A_i}\neq \hat{U}^{B_i}\hat{V}_1\hat{U}^{A_i} $  i.e., in that frame while the order of operations is common, when they take place relative to periods of free dynamics, and thus the overll evolution,  depends on the control. Crucially, in this case temporal order among events cannot be even defined as changes depending on the reference frame.
A further discussion of the role played by free evolution in this protocol is presented in Supplementary Note 4.

In fact, if free evolution and the applied operations do not commute, even simpler scenarios can illustrate the issue.  
Consider that one of the operations is trivial, say $\hat U^B=\hat{\mathbb I}$, and  so in fact only \textit{one} operation is applied. The non-commutativity between $\hat U_A$ and $\hat V=\hat V_0=\hat V_1$ (we are in the reference frame of the cavity)  would again result in different final states depending on when $\hat U^A$ is applied relative to $\hat V$. This would again lead to an entangled final state in a two-wing scenario, even though in this case there is even no time order of events to speak of.

Finally, we note that in all quantum switch scenarios, including the entangled switch, there is an assumption that each local operation is performed `only once'. This condition is, however, naturally satisfied in the relativistic implementations such as ours, as each operation is performed at a specific time of a local clock.

\section*{Conclusion}\label{sec:discussion}
In this paper, we constructed a non-gravitational scenario where accelerating particles, interacting with quantum fields,
according to their own internal clock degrees of freedom, can lead to a violation of the temporal Bell inequalities
analogous to the gravitational case.
In the ``Methods'' section we introduced the formalism required to reproduce the gravitational protocol using special relativistic time dilation. We defined the kinematics of all particles involved in the protocol and the appropriate coupling between them and the quantum field. In the ``Results'' section, we proposed the full protocol, that demonstrated indefinite causal order in a fully operational
setting. We described the procedure than would lead us to the violation of Bell's inequalities for the proposed system. We found that a violation of Bell inequalities occurs also when the entangled events are space-like what we interpreted as an ambiguity in the signature of indefinite temporal order.
We finally found that this surprising conclusion is the result of failure of the additional assumption, requiring that target systems have no other evolution except, the one governed by unitaries applied in a specific time order. A detailed discussion of this problem we presented in the ``Ambiguity in the signature of indefinite temporal order'' section of our work.

We have focused so far on a particular realisation of the entangled switch protocol -- with two-level `detectors' and cavity-confined quantum field modes as targets, and the position DoFs of the detectors as the control. However, our main result and its explanation  applies to any physical realisation of the protocol. Indeed, we have shown that entanglement can be generated for spacelike separated operations and identified that this is due to free evolution of the targets. For any physical implementation of the protocol, if the applied operations do not commute with the free evolution,  the final state will in general be entangled regardless of the commutation relations between the operations themselves, and thus also regardless of their temporal order.  Entanglement can arise even if only one operation is applied, as discussed in the previous section.

It is of course possible to identify implementations where the assumption of no free evolution of the targets does hold
(e.g.~with polarisation or angular momentum DoFs of photons) or where the operations are performed within a degenerate subspace of the Hamiltonian of the targets (and thus commute with free evolution), so that only non-commutativity between the two local operations is relevant. Such strategies would indeed lead to an entangled final state only if the local operations are  timelike and applied in a non-classical order. However, identifying such implementations does require a theoretical description of the states and dynamics of the involved systems. In other words, one needs theory-dependent assumptions to interpret a violation of the final Bell inequalities as a signature of indefinite temporal order.

The observations above do not depend on the special-relativistic setting studied in this work and hold equally for gravitational protocols. Indeed, nothing in the argument depends on how the time dilation of the clocks is achieved. For example, the position of a massive body can determine -- through gravitational time dilation -- the time at which a single operation takes place (relative to some mass-independent coordinates). This can lead to the same situation as described earlier: creation of entanglement (in a two-wing scenario) even with a single operation per wing. Again, theory-dependent assumptions would be required to ensure that entanglement can only arise as a result of an indefinite order of events.

Being able to describe and experimentally test entangled temporal order (i.e.~even working fully within quantum mechanics) is of interest in its own right. Our result, however, opens the question whether it is possible to formulate a stronger,  theory independent, test of temporal order.
The insight from the present study is that it is problematic to separate out the effect of the free dynamics of the system from that of the local operations. A possible avenue to circumvent this is to consider more general operations than the fixed unitaries discussed thus far: they can involve a measurement of the system, producing a classical variable as outcome. Furthermore, a setting variable for each party can model a choice among different operations. In this fashion, one can consider directly the causal relations between parties -- understood operationally as correlations between settings and outcomes -- without relying on a theory-dependent description of the transformations. We leave further investigation of this possibility to future work.

\section{acknowledgements}
K.D. is financially supported by the (Polish) National Science Center Grant 2021/41/N/ST2/01901.
F.C.~acknowledges support through the Australian Research Council (ARC) DECRA grant DE170100712, M.Z.~acknowledges support through ARC Future Fellowship grant FT210100675, F.C.and M.Z. acknowledge support through ARC Centre of Excellence EQuS CE170100009. The University of Queensland (UQ) acknowledges the Traditional Owners and their custodianship of the lands on which UQ operates.
 


\bibliographystyle{nature}
\bibliography{cavity}



\onecolumngrid

\title{Supplementary Information: Indefinite temporal order -- cavity model
}

\author{XXX}
\affiliation{}

\author{XXX}
\affiliation{}

\author{XXX}
\affiliation{}

\author{XXX}
\affiliation{}

\date{\today}

\maketitle

\onecolumngrid
 
\newpage
\section*{Supplementary Note 1: Second order of the Dyson series}\label{2orderDyson}

In this section, We will show that second-order Dyson expansion does not affect on $\ket{\Phi_R}$ and $\ket{\Phi_L}$. 

\medskip

One can ask the question why \eqref{eqn:PhiRLscalarapprox} depends on $\lambda^2$. Previously we limited our calculation only to the first-order expansion of the Dyson series. We have to verify that second-order expansion does not produce lambda square terms too. Otherwise, there is a possibility that results from \eqref{eqn:PhiRLscalarapprox} will cancel out with these additional terms. Let us write the general evolution operator in the following form:
\begin{gather}
\hat{U}
=
\mathds{1}-i
\lambda\int_{-\infty}^{\infty} 
\mathrm{d} t
~\chi_{d}(t)
\left(e^{i\Omega t} \hat{\sigma}^{+}+e^{-i\Omega t} \hat{\sigma}^{-}\right)
\sum_{n}\left(\hat{a}_n^{\dagger} u_{n}(x_d)e^{i\omega_n t}+
\hat{a}_n u_{n}(x_d)e^{-i\omega_n t}\right)
\\
-
\lambda^2\int_{-\infty}^{\infty} 
\mathrm{d} t_2
\int_{-\infty}^{t_2} \mathrm{d} t_1
~\chi_{d}(t_2)
\chi_{d}(t_1)
\left(e^{i\Omega t_2} \hat{\sigma}^{+}+e^{-i\Omega t_2} \hat{\sigma}^{-}\right)
\left(e^{i\Omega t_1} \hat{\sigma}^{+}+e^{-i\Omega t_1} \hat{\sigma}^{-}\right)
\nonumber
\\
\times\sum_{n}\left(\hat{a}_n^{\dagger} u_{n}(x_d)e^{i\omega_n t_2}+
\hat{a}_n u_{n}(x_d)e^{-i\omega_n t_2}\right)
\sum_{m}\left(\hat{a}_m^{\dagger} u_{m}(x_d)e^{i\omega_m t_1}+
\hat{a}_m u_{m}(x_d)e^{-i\omega_m t_1}\right)\nonumber,
\end{gather} 
where $x_d$ is a position of a detector and $\hat{\sigma}^{\pm}$ acts on the internal state of this detector. In our case we have two detectors and two Hilbert spaces of internal degrees of freedom.  We can simplify the notation to write operators of the evolution as:
\begin{gather}
   \hat{U_1}
   =
\mathds{1}-i\lambda\sum_{n}\left(\hat{U}^{+}_{1n}\hat{\sigma}^{+}_{1}+\hat{U}^{-}_{1n}\hat{\sigma}^{-}_{1}\right)
-
\lambda^2
\sum_{n,m}
\left(
\hat{U}^{++}_{1nm}
\hat{\sigma}^{+}_{1}\hat{\sigma}^{+}_{1}
+
\hat{U}^{+-}_{1nm}
\hat{\sigma}^{+}_{1}\hat{\sigma}^{-}_{1}
+
\hat{U}^{-+}_{1nm}
\hat{\sigma}^{-}_{1}\hat{\sigma}^{+}_{1}
+
\hat{U}^{--}_{1nm}
\hat{\sigma}^{-}_{1}\hat{\sigma}^{-}_{1}
\right),
\\
   \hat{U_2}
   =
\mathds{1}-i\lambda\sum_{n}\left(\hat{U}^{+}_{2n}\hat{\sigma}^{+}_{2}+\hat{U}^{-}_{2n}\hat{\sigma}^{-}_{2}\right)
-
\lambda^2
\sum_{n,m}
\left(
\hat{U}^{++}_{2nm}
\hat{\sigma}^{+}_{2}\hat{\sigma}^{+}_{2}
+
\hat{U}^{+-}_{2nm}
\hat{\sigma}^{+}_{2}\hat{\sigma}^{-}_{2}
+
\hat{U}^{-+}_{2nm}
\hat{\sigma}^{-}_{2}\hat{\sigma}^{+}_{2}
+
\hat{U}^{--}_{2nm}
\hat{\sigma}^{-}_{2}\hat{\sigma}^{-}_{2}
\right),
\end{gather}
where $U^{\pm}_{1n}$, $U^{\pm}_{2n}$ are operators from the first-order order expansion and $U^{\pm\pm}_{1nm}$ and $U^{\pm\pm}_{2nm}$ are operators from the second-order expansion of the Dyson series. $\hat{\sigma}^{\pm}_{1}$ and $\hat{\sigma}^{\pm}_{2}$ are operators $\hat{\sigma}^{\pm}$ acting on the internal space of the first or the second detector respectivly. To find contributions proportional to the $\lambda^2$ to the value of \eqref{eqn:PhiRLscalarapprox}, we have to find new terms in the $\ket{\psi_R}$ and $\ket{\psi_L}$ proportional to $\ket{ge}$ or $\ket{eg}$. Knowing that $\ket{\psi_R}=\hat{U}_1\hat{U}_2\ket{gg}\ket{0}$, $\ket{\psi_L}=\hat{U}_2\hat{U}_1\ket{gg}\ket{0}$, $\hat{\sigma}^{-}\ket{g}=0$ and $\hat{\sigma}^{+}\ket{e}=0$ we can write:
\begin{gather}
\ket{\psi_R}=
\left[
\mathds{1}-i\lambda\sum_{n}\hat{U}^{+}_{1n}\hat{\sigma}^{+}_{1}
-
\lambda^2
\sum_{n,m}
\hat{U}^{-+}_{1nm}
\hat{\sigma}^{-}_{1}\hat{\sigma}^{+}_{1}
\right]
\left[
\mathds{1}-i\lambda\sum_{n}\hat{U}^{+}_{2n}\hat{\sigma}^{+}_{2}
-
\lambda^2
\sum_{n,m}
\hat{U}^{-+}_{2nm}
\hat{\sigma}^{-}_{2}\hat{\sigma}^{+}_{2}
\right]\ket{gg}\ket{0},
\\
\ket{\psi_L}=
\left[
\mathds{1}-i\lambda\sum_{n}\hat{U}^{+}_{2n}\hat{\sigma}^{+}_{2}
-
\lambda^2
\sum_{n,m}
\hat{U}^{-+}_{2nm}
\hat{\sigma}^{-}_{2}\hat{\sigma}^{+}_{2}
\right]
\left[
\mathds{1}-i\lambda\sum_{n}\hat{U}^{+}_{1n}\hat{\sigma}^{+}_{1}
-
\lambda^2
\sum_{n,m}
\hat{U}^{-+}_{1nm}
\hat{\sigma}^{-}_{1}\hat{\sigma}^{+}_{1}
\right]\ket{gg}\ket{0}.
\end{gather}
We can observe that second-order contributions from the same detector do not change the internal state, i.e.~$\hat{\sigma}^{-}\hat{\sigma}^{+}\ket{g}=\ket{g}$, while the product of first order terms from both detectors gives $\hat{\sigma}_{1}^{+}\hat{\sigma}_{2}^{+}\ket{gg}=\hat{\sigma}_{2}^{+}\hat{\sigma}_{1}^{+}\ket{gg}=\ket{ee}$. Thus, the second order terms do not have support on the subspace spanned by the states $\ket{ge}, \ket{eg}$. Note that the states $\ket{ge}$ or $\ket{eg}$  appear in this expansion at order  $\lambda^3$ or higher.  This means that second-order terms from the Dyson series do not affect the states $\ket{\Phi_R}$ and $\ket{\Phi_L}$ in our approximation.

\section*{Supplementary Note 2: Details of calculations of the final state}\label{Details}
Using the form of evolution operator \eqref{eqn:U} we can find that:
\begin{gather}
   \ket{\psi_R}=\hat{U}_1\hat{U}_2\ket{g}\ket{g}\ket{0}=
    \ket{g}\ket{g}\ket{0}
    -i\lambda\int\mathrm{d}t_2\chi_{2R}(t_2)\sum_{k}\frac{1}{\sqrt{\omega_k L}}e^{i(\omega_k+\Omega)t_2}\sin(\frac{k\pi}{L}x_2)\ket{g}\ket{e}\hat{a}_k^{\dagger}\ket{0}\nonumber\\
    -i\lambda\int\mathrm{d}t_1\chi_{1R}(t_1)\sum_{k}\frac{1}{\sqrt{\omega_k L}}e^{i(\omega_k+\Omega)t_1}\sin(\frac{k\pi}{L}x_1)\ket{e}\ket{g}\hat{a}_k^{\dagger}\ket{0}
    +\mathcal{O}(\lambda^2),
\end{gather}
where: $\chi_{1R}$--switching function for the first detector in the case that right detector interacts before the left one,  $\chi_{2R}$--switching function for the second detector in the case that right detector interacts earlier. We assume that the interaction starts and ends rapidly, so that $\chi_{2R}(t)=1$ for $t\in(0,T)$ and $\chi_{2R}(t)=0$ for any other time. Similarly, $\chi_{1R}(t)=1$ for $t\in(\Delta\tau,\Delta\tau+T)$ and  $\chi_{1R}(t)=0$ for any other time. We can proceed with the same calculation for $\ket{\psi_L}$:
\begin{gather}
   \ket{\psi_L}= \hat{U}_2\hat{U}_1\ket{g}\ket{g}\ket{0}=
    \ket{g}\ket{g}\ket{0}
    -i\lambda\int\mathrm{d}t_2\chi_{2L}(t_2)\sum_{k}\frac{1}{\sqrt{\omega_k L}}e^{i(\omega_k+\Omega)t_2}\sin(\frac{k\pi}{L}x_2)\ket{g}\ket{e}\hat{a}_k^{\dagger}\ket{0}\nonumber\\
    -i\lambda\int\mathrm{d}t_1\chi_{1L}(t_1)\sum_{k}\frac{1}{\sqrt{\omega_k L}}e^{i(\omega_k+\Omega)t_1}\sin(\frac{k\pi}{L}x_1)\ket{e}\ket{g}\hat{a}_k^{\dagger}\ket{0}
    +\mathcal{O}(\lambda^2),
\end{gather}
where: $\chi_{1L}$--switching function for the first detector in the case that right detector interacts before the left one,  $\chi_{2L}$--switching function for the second detector in the case that right detector interacts earlier. In this case we have that $\chi_{1L}(t)=1$ for $t\in(0,T)$ and $\chi_{1L}(t)=0$ for any other time. Similarly, $\chi_{2L}(t)=1$ for $t\in(\Delta\tau,\Delta\tau+T)$ and  $\chi_{2L}(t)=0$ for any other time. It worth noticing that $\chi_{1L}=\chi_{2R}$ and $\chi_{2L}=\chi_{1R}$. After using this property describing relation between switching functions, we have:
\begin{gather}
    \ket{\psi_R}=\hat{U}_1\hat{U}_2\ket{g}\ket{g}\ket{0}=
    \ket{g}\ket{g}\ket{0}
    -i\lambda\int\mathrm{d}t_2\chi_{2R}(t_2)\sum_{k}\frac{1}{\sqrt{\omega_k L}}e^{i(\omega_k+\Omega)t_2}\sin(\frac{k\pi}{L}x_2)\ket{g}\ket{e}\hat{a}_k^{\dagger}\ket{0}\nonumber\\
    -i\lambda\int\mathrm{d}t_1\chi_{1R}(t_1)\sum_{k}\frac{1}{\sqrt{\omega_k L}}e^{i(\omega_k+\Omega)t_1}\sin(\frac{k\pi}{L}x_1)\ket{e}\ket{g}\hat{a}_k^{\dagger}\ket{0}
    +\mathcal{O}(\lambda^2),
\end{gather}
\begin{gather}
    \ket{\psi_L}=\hat{U}_2\hat{U}_1\ket{g}\ket{g}\ket{0}=
    \ket{g}\ket{g}\ket{0}
    -i\lambda\int\mathrm{d}t_2\chi_{2R}(t_2)\sum_{k}\frac{1}{\sqrt{\omega_k L}}e^{i(\omega_k+\Omega)t_2}\sin(\frac{k\pi}{L}x_1)\ket{e}\ket{g}\hat{a}_k^{\dagger}\ket{0}
    \nonumber
    \\
    -i\lambda\int\mathrm{d}t_1\chi_{1R}(t_1)\sum_{k}\frac{1}{\sqrt{\omega_k L}}e^{i(\omega_k+\Omega)t_1}\sin(\frac{k\pi}{L}x_2)\ket{g}\ket{e}\hat{a}_k^{\dagger}\ket{0} +\mathcal{O}(\lambda^2).
\end{gather}
For simplicity of further calculation, let us introduce the following notation: 
\begin{gather}
\ket{\psi_R}=\hat{U}_1\hat{U}_2\ket{g}\ket{g}\ket{0}=\ket{gg}\ket{0}+\ket{ge}\ket{\phi_{ge}^R}+\ket{eg}\ket{\phi_{eg}^R}+\mathcal{O}(\lambda^2),
\label{eqn:psiR_A}
\end{gather}
\begin{gather}
\ket{\psi_L}=\hat{U}_2\hat{U}_1\ket{g}\ket{g}\ket{0}=\ket{gg}\ket{0}+\ket{ge}\ket{\phi_{ge}^L}+\ket{eg}\ket{\phi_{eg}^L}+\mathcal{O}(\lambda^2),
\label{eqn:psiL_A}
\end{gather}
where: $\ket{\phi_{ge}^L}$--a state of a field when the left detector interacts first but the right detector is excited, $\ket{\phi_{eg}^L}$--a state of a field when the left detector interacts first and the left detector is excited, $\ket{\phi_{ge}^R}$--a state of a field when the right detector interacts first and the right detector is excited, $\ket{\phi_{eg}^R}$--a state of a field when the right detector interacts first but the left detector is excited as a consequence of the interaction between atoms and the field.

\section*{Supplementary Note 3: Method of finding appropriate parameters}\label{parameters}
In this section we will find an appropriate set of parameters that orthogonalize $\ket{\Phi_R}$ and $\ket{\Phi_L}$ vectors.

\begin{align}
    \ket{\Phi_R}
    &=
    -i\lambda
    \sum_{k}\frac{1}{\sqrt{\omega_k L}}
    \int\mathrm{d}t
    \left(
    \chi_{2R}(t)\sin(\frac{k\pi}{L}x_2)
    +
    \chi_{1R}(t)\sin(\frac{k\pi}{L}x_1)
    \right)
    e^{i(\omega_k+\Omega)t}
    \hat{a}_k^{\dagger}\ket{0}
    \nonumber
    \\
    &=
    -i\lambda
    \sum_{k}
    \int\mathrm{d}t
    \left(
    \chi_{2R}(t)u_k(x_2)
    +
    \chi_{1R}(t)u_k(x_1)
    \right)
    e^{i(\omega_k+\Omega)t}
    \hat{a}_k^{\dagger}\ket{0}
    \nonumber
    \\
    &=
    -\lambda
    \sum_{k}
    \frac{e^{iT(\omega_k+\Omega)}-1}{\omega_k+\Omega}\left(e^{i\Delta\tau(\omega_k+\Omega)}u_k(x_1)+u_k(x_2)\right)
    \hat{a}_k^{\dagger}\ket{0}
    \\
    \ket{\Phi_L}
    &=
    -i\lambda
    \sum_{k}\frac{1}{\sqrt{\omega_k L}}
    \int\mathrm{d}t
    \left(
    \chi_{2R}(t)\sin(\frac{k\pi}{L}x_1)
    +
    \chi_{1R}(t)\sin(\frac{k\pi}{L}x_2)
    \right)
    e^{i(\omega_k+\Omega)t}
    \hat{a}_k^{\dagger}\ket{0}
    \nonumber
    \\
    &=
    -\lambda
    \sum_{k}
    \frac{e^{iT(\omega_k+\Omega)}-1}{\omega_k+\Omega}\left(e^{i\Delta\tau(\omega_k+\Omega)}u_k(x_2)+u_k(x_1)\right)
    \hat{a}_k^{\dagger}\ket{0}
\end{align}

\begin{figure*}[b]
\centering
    \includegraphics[width=0.8\linewidth]{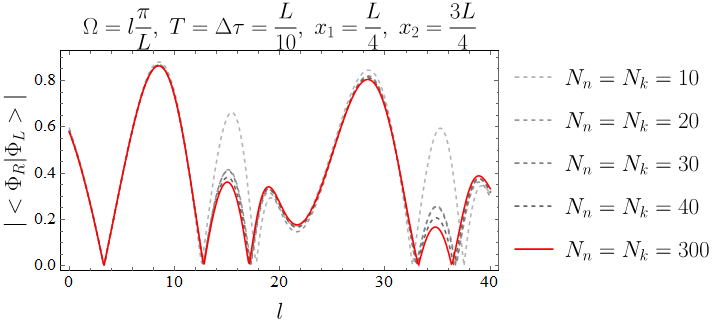}
    \caption{{\bf The analysis of an orthogonality between $\ket{\Phi_L}$ and $\ket{\Phi_L}$.}The absolute value of the scalar product as a function of the energy gap of the detector $\Omega$.}
        \label{fig:ABSomega}
\end{figure*}

Where the form of the state $\ket{\Phi_L}$ we get by changing $x_1\longleftrightarrow x_2$. Now it is easy to see that the scalar product can be written as:
\begin{align}
    \braket{\Phi_R}{\Phi_L}
    &\cong
    \lambda^2
    \sum_{k}
    \frac{\abs{e^{iT(\omega_k+\Omega)}-1}^2}{\left(\omega_k+\Omega\right)^2}\left(e^{-i\Delta\tau(\omega_k+\Omega)}u_k(x_1)+u_k(x_2)\right)
    \left(e^{i\Delta\tau(\omega_k+\Omega)}u_k(x_2)+u_k(x_1)\right)
    \nonumber
    \\
    &\cong
    2\lambda^2
    \sum_{k}
    \frac{e^{-i\Delta\tau(\omega_k+\Omega)}}{(\omega_k+\Omega)^2}
    \left[
    1-\cos\left(T(\omega_k+\Omega)\right)
    \right]
    \left(
    u_k(x_1)
    +
    e^{i\Delta\tau(\omega_k+\Omega)}
    u_k(x_2)
    \right)^2
    \label{eqn:PhiRLscalarapprox}
\end{align}
We have to remember that states $\ket{\Phi_L}$ and $\ket{\Phi_R}$ were not properly normalized so now we can find the norm:
\begin{align}
    \norm{\ket{\Phi_R}}&=\sqrt{\braket{\Phi_R}{\Phi_R}}
    =
    \sqrt{
    \lambda^2
    \sum_{k}
    \frac{\abs{e^{iT(\omega_k+\Omega)}-1}^2}{\left(\omega_k+\Omega\right)^2}
    \abs{e^{i\Delta\tau(\omega_k+\Omega)}u_k(x_1)+u_k(x_2)}^2
    }
    \nonumber
    \\
    &=
    2\lambda
    \sqrt{
    \sum_{k}
    \frac{\sin^2\frac{T(\omega_k+\Omega)}{2}}{(\omega_k+\Omega)^2}
    \left[
    u_k(x_1)^2
    +
    2u_k(x_1)u_k(x_2)\cos\left(\Delta\tau(\omega_k+\Omega)\right)
    +u_k(x_2)^2
    \right]
    }
    \\
    \norm{\ket{\Phi_L}}&=\sqrt{\braket{\Phi_L}{\Phi_L}}=
    2\lambda
    \sqrt{
    \sum_{k}
    \frac{\sin^2\frac{T(\omega_k+\Omega)}{2}}{(\omega_k+\Omega)^2}
    \left[
    u_k(x_2)^2
    +
    2u_k(x_2)u_k(x_1)\cos\left(\Delta\tau(\omega_k+\Omega)\right)
    +u_k(x_1)^2
    \right]
    }
\end{align}
We can notice that $\norm{\ket{\Phi_R}}=\norm{\ket{\Phi_L}}$ and finally the scalar product has the following form:
\begin{align}
\braket{\Phi_R}{\Phi_L}
=
\frac{
\braket{\Phi_R}{\Phi_L}
}
{
\norm{\ket{\Phi_R}}\norm{\ket{\Phi_L}}
}
=
\frac{
    \sum_{k}
    \frac{e^{-i\Delta\tau(\omega_k+\Omega)}}{(\omega_k+\Omega)^2}
    \left[
    1-\cos\left(T(\omega_k+\Omega)\right)
    \right]
    \left(
    u_k(x_1)
    +
    e^{i\Delta\tau(\omega_k+\Omega)}
    u_k(x_2)
    \right)^2
}{
2
    \sum_{n}
    \frac{\sin^2\frac{T(\omega_n+\Omega)}{2}}{(\omega_n+\Omega)^2}
    \left[
    u_n(x_2)^2
    +
    2u_n(x_2)u_n(x_1)\cos\left(\Delta\tau(\omega_n+\Omega)\right)
    +u_n(x_1)^2
    \right]
}
\end{align}
This function can be estimated as a finite sum. Let us denote the upper limits of these two sums as $N_n$ for the sum over $n$ and $N_k$ for the sum over $k$.

\begin{figure*}
    \centering
    \includegraphics[width=0.3\columnwidth]{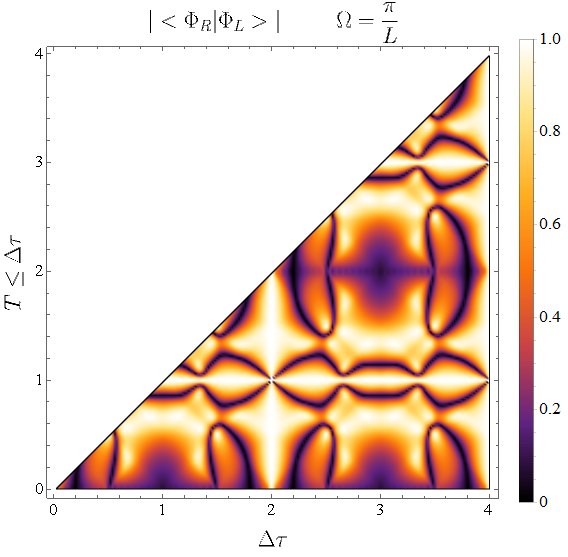}\quad
    \includegraphics[width=0.3\columnwidth]{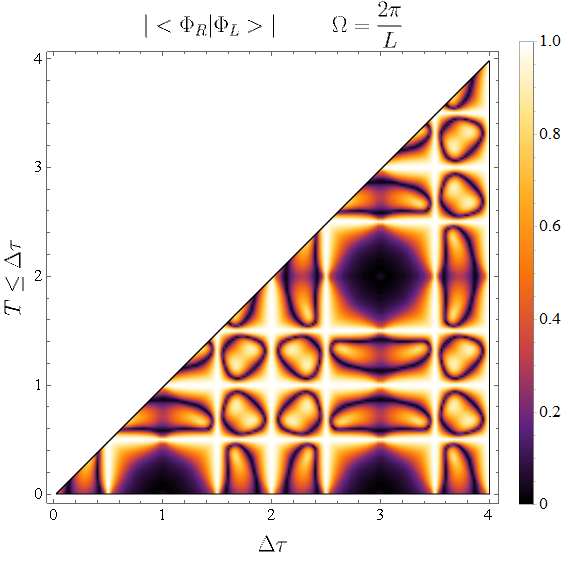}\quad
    \includegraphics[width=0.3\columnwidth]{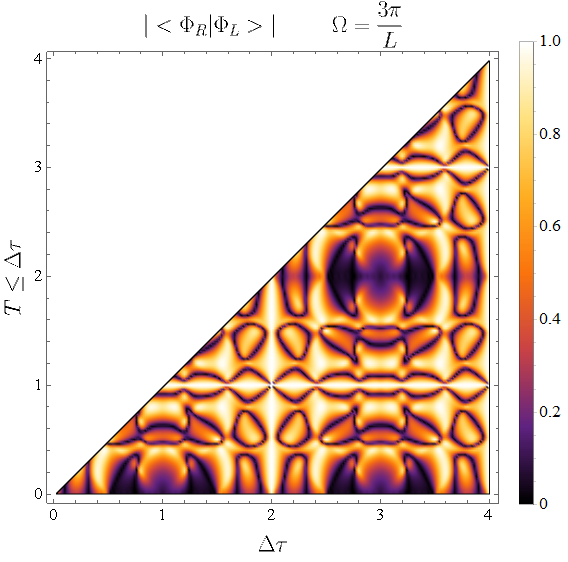}
    \caption{{\bf The analysis of an orthogonality between $\ket{\Phi_L}$ and $\ket{\Phi_L}$ for different parameters describing interaction.}\\An absolute value of the scalar product for two detectors standing at $x_1=L/4$ and $x_2=3L/4$. The energy gap $\Omega$ is chosen as one of the cavity frequencies. Scalar product approximated as a finite sum of $30$ modes i.e. $N_k=N_n=30$.}
    \label{fig:ABStauT}
\end{figure*}

Fig.~\ref{fig:ABSomega} shows the absolute value of the scalar product as a function of the energy gap $\Omega$ plotted for different numbers of modes $N_n$ and $N_k$. We can see that we can produce entanglement for very short times $\Delta\tau=T=L/10$. Thus, it proves that there is an entanglement between two cavities also for spacelike separated events of interaction.

Fig.~\ref{fig:ABSomega} look quite random. Let us analyze the scalar product for the case of two detectors standing in the positions $x_1=L/4$ and $x_2=3L/4$. Fig.~\ref{fig:ABStauT} show the absolute value of the scalar product for different parameters $\Delta\tau$ and $T\leq\Delta\tau$ for different values of the energy gap $\Omega$. We can notice that there are many parameters minimizing the scalar product between two states. 

Based on Fig.~\ref{fig:ABStauT} we can conjecture that point $\left(\Delta\tau,T\right)=\left(3L,2L\right)$ is a good candidate for the orthogonalization of the states $\ket{\Phi_R}$ and $\ket{\Phi_L}$. To verify this hypothesis let us consider the following calculation: 
Let $L=1$, $x_1=1/4$, $x_2=3/4$, $\Omega=\pi$, $\Delta\tau=3$, $T=2+\epsilon$, where $\epsilon\in\mathbb{R}_+$ is a small parameter.  Then
\begin{align}
\braket{\Phi_R}{\Phi_L}
=
-
\frac{
\sum_{k}\frac{e^{-3ik\pi}}{k(k+1)^2}
\left[
\cos\left((1+k)(2+\epsilon)\pi\right)-1
\right]
\left(
\sin\frac{k\pi}{4}
-e^{3ik\pi}
\sin\frac{3k\pi}{4}
\right)^2
}{
2\sum_{n}
\frac{\sin^2\frac{(n+1)\pi\epsilon}{2}}{2n(n+1)^2}
\left[
(1+2(-1)^n)\cos\frac{n\pi}{2}+\cos\frac{3n\pi}{2}-4
\right]
}.
\end{align}
Each of these sums can be done analytically for arbitrary $\epsilon$. Then we expand the numerator and denominator around $\epsilon=0$, to get:
\begin{align}
\abs{
\braket{\Phi_R}{\Phi_L}
}
=
\abs{
\frac{
\frac{1}{2\pi}\left(2\log20\log(1-i)-\log(1+i)\right)\epsilon^2+\mathcal{O}(\epsilon^3)
}{
\frac{1}{6\pi}\left(9+\log8-6\log\pi-6\log\epsilon\right)\epsilon^2
+\mathcal{O}(\epsilon^3)
}
}
\approx
\frac{
\log8
}{
9+\log8-6\log\pi\epsilon
}.
\end{align}
And we see that:
\begin{align}
\lim_{\epsilon\to0^+}\abs{\braket{\Phi_R}{\Phi_L}}
=0
\end{align}

\section*{Supplementary Note 4: Indefinite order for spacelike separation of events}\label{spacelike}

Results from the main text of the ``Entanglement of the final state'' section show that also for spacelike separation of events of interactions Bell's inequality for temporal order can be violated. This statement seems paradoxical 
because for spacelike separation of events we cannot define their  temporal order - it would depend on the reference frame. Here we look more closely at the compatibility of this fact with the locality of evolution.
\medskip

\begin{figure*}[h]
\centering
    \includegraphics[width=0.9\linewidth]{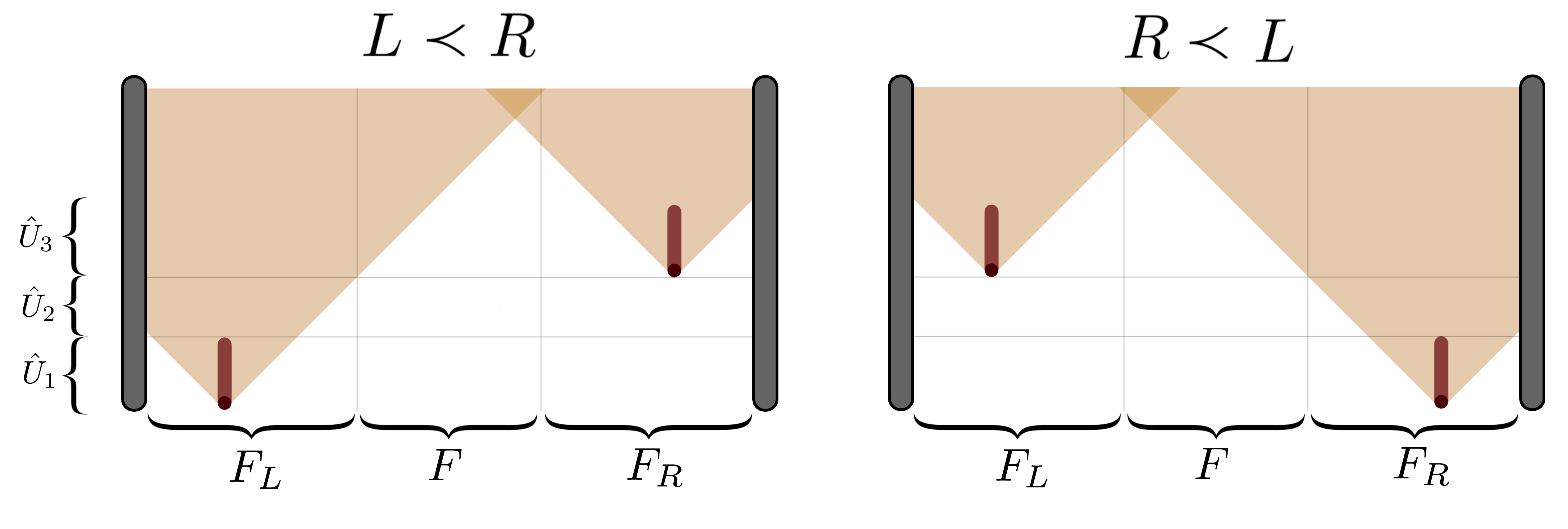}
        \caption{{\bf Spacetime diagram of a cavity interacting with two detectors.} Operators $\hat{U}_1$ and $\hat{U}_3$ describes the evolution of the system according to the interaction, and $\hat{U}_2$ is an operator of the free evolution that occurs between interactions. $F_L$ and $F_R$ are parts of the cavity that can interacts with the detector in some finite time. $F$ is a middle segment of the cavity that evolves only due to the free evolution operator between interactions.}
        \label{fig:AppendixD}
\end{figure*}
Let us consider the cavity and two molecules placed near its boundaries. We know that entnaglement at spacelike separation appears for a short time of interaction and small time dilation (see Fig.~\ref{fig:grid}). By definition, in such scenarios information about the interaction with the field can not be transferred from the first to the second detector before the latter interacts with the field. To notice the problem with this situation we can split the whole cavity into three parts (see Fig. \ref{fig:AppendixD}). Let us denote these as: $F_L$, $F$, and $F_R$. The field $F_L$ refers only to the part of the cavity nearby the left detector and similarly $F_R$ is describing just the right part of the cavity. The field $F$ refers to the middle of the whole cavity. The lengths of each part are chosen to make sure that information about interaction with the left/right molecule can be localized only within the left/right part of the cavity. Because of this division of the cavity into three parts, we can focus on two alternative orders of events. Let us focus on the situation when the left molecule interacts first ($L\prec R$). The whole evolution $\hat{U}$ of the system containing two molecules and three parts of the cavity can be presented as $\hat{U}=\hat{U}_3~\hat{U}_2~\hat{U}_1$, where: 
\begin{itemize}
    \item $\hat{U}_1$ -- interaction of the left molecule with $F_L$ and free evolution of remaining parts.
    \item $\hat{U}_2$ -- free evolution of the whole cavity.
    \item $\hat{U}_3$ -- interaction of the right molecule with the $F_R$ and free evolution of remaining parts.
\end{itemize}
Alternatively, for the right molecule interacting first  ($R\prec L$), we would have:
\begin{itemize}
    \item $\hat{U}_1$ -- interaction of the right molecule with the $F_R$ and free evolution of remaining parts.
    \item $\hat{U}_2$ -- free evolution of the whole cavity.
    \item $\hat{U}_3$ -- interaction of the left molecule with the $F_L$ and free evolution of remaining parts.
\end{itemize}
We can notice that these two scenarios differ by not only the order of events but also by the free evolution parts. It means that for such a case we can not argue that violation of Bell's inequalities imply indefinite order of events because we violate one of the assumptions. We are considering a scenario in which not only the order of events are different but also there is an additional part of the evolution of the system that causes a different final state for each order.

\end{document}